\RequirePackage{fix-cm}

\documentclass[smallextended]{svjour3} 
\smartqed
\usepackage{cite}
\usepackage{amsmath,amssymb,amsfonts}
\usepackage{algorithmic}
\usepackage{graphicx}
\usepackage{textcomp}
\usepackage{longtable}
\usepackage{todonotes}
\usepackage{multirow}
\usepackage{multicol}
\usepackage[authoryear, round]{natbib}
\usepackage[many]{tcolorbox}
\usepackage{colortbl}
\usepackage{float}
\usepackage{placeins}
\usepackage[hyphens]{url}
\usepackage{hyperref}
\usepackage{enumitem}
\usepackage{soul}
\usepackage{balance}

\tcbset{
    sharp corners,
    colback = white,
}
\renewcommand\st[1]{}

\newtcolorbox{boxD}{
    colback = {white},
    colframe = {black},
    boxrule = 0pt,
    toprule = 1.5pt,
    bottomrule = 1.5pt,
    left=3pt,right=3pt, top=2pt,bottom=2pt
}

\begin{document}

\title{Plug it and Play on Logs: A Configuration-Free Statistic-Based Log Parser}

\author{Qiaolin Qin         \and
        Xingfang Wu         \and
        Heng Li         \and
        Ettore Merlo 
}

\institute{Qiaolin Qin, Xingfang Wu, Heng Li, Ettore Merlo \at
              Department of Computer Engineering and Software Engineering \\
              Polytechnique Montreal\\
              Montreal, QC, Canada\\
              \email{\{qiaolin.qin, xingfang.wu, heng.li, ettore.merlo\}@polymtl.ca}        
}

\date{Received: date / Accepted: date}



\maketitle

\begin{abstract}
\label{abstract}
Log parsing is an essential task in log analysis, and many tools have been designed to accomplish it. Existing log parsers can be categorized into statistic-based and semantic-based approaches. In comparison to semantic-based parsers, existing statistic-based parsers tend to be more efficient, require lower computational costs, and be more privacy-preserving thanks to on-premise deployment, but often fall short in their accuracy (e.g., grouping or parsing accuracy) and generalizability. Therefore, it became a common belief that statistic-based parsers cannot be as effective as semantic-based parsers since the latter could take advantage of external knowledge supported by pretrained language models. Our work, however, challenges this belief with a novel statistic-based parser, PIPLUP. PIPLUP eliminates the pre-assumption of the position of constant tokens for log grouping and relies on data-insensitive parameters to overcome the generalizability challenge, 
allowing ``plug and play'' on given log files. According to our experiments on an open-sourced large log dataset, PIPLUP shows promising accuracy and generalizability with the data-insensitive default parameter set. PIPLUP not only outperforms the state-of-the-art statistic-based log parsers, Drain and its variants, but also obtains a competitive performance compared to the best unsupervised semantic-based log parser (i.e., LUNAR). Further, PIPLUP exhibits low time consumption without GPU acceleration and external API usage; our simple, efficient, and effective approach makes it more practical in real-world adoptions, especially when costs and privacy are of major concerns. 
\end{abstract}


\section{Introduction}
\label{sec:introduction}
Log messages are generated by logging statements in the source code which typically consist of a log template and parameters that record dynamic information (e.g., system status or variable values). Log messages capture rich runtime information of software systems, which is vital for understanding software behaviors and can help enhance the software maintenance process. Log parsing is a common approach to compress the high volume of log files into event templates with semantic and structural features. Using the templates, researchers and practitioners can better analyze and maintain the systems. For example, prior work leveraged log parsing for log-based anomaly detection~\citep{kuttal2011history, gadler2017mining, fu2009execution, he2016experience, du2017deeplog, chen2021experience, wu2023effectiveness}, root cause analysis~\citep{lu2017log, yuan2012characterizing, fu2013contextual}, or performance regression detection~\citep{liao2020using, chow2014mystery, nagaraj2012structured}. Hence, in the field of software engineering, log analysis is a highly important topic. 
%


Many studies on automatic log parsing have been carried out through the years. Log parsers can be broadly categorized into statistic-based~\citep{dai2020logram, nagappan2010abstracting, vaarandi2015logcluster,fu2009execution, hamooni2016logmine, mizutani2013incremental, shima2016length, tang2011logsig,du2016spell, he2017drain, makanju2009clustering, messaoudi2018search, yu2023brain} and semantic-based ones~\citep{le2023log, li2023did, liu2022uniparser,jiang2024lilac,ma2024librelog,xiao2024free,huang2025no}. Statistics-based parsers leverage heuristic rules for parsing in unsupervised fashion, while semantic-based parsers usually use deep learning models (e.g., LSTM and LLMs) to encode and interpret log messages, group the logs, and extract event templates based on the semantic information. 
\citet{jiang2024large} conducted a large-scale study that discusses the existing log parsers' pros and cons in terms of their efficiency (i.e., time consumption) and effectiveness (i.e., log grouping and parsing quality). 
According to the experimental results, Jiang \textit{et al.} discovered that statistic-based parsers tend to have good log grouping performance and are time-efficient, but their parsing accuracies are less satisfactory. In contrast, while semantic-based parsers can achieve high parsing accuracy, they often overlook global information, leading to low grouping accuracy and requiring more processing time and computational resources (e.g., GPUs). Further, the usage of black-box models (e.g., ChatGPT) raises privacy concerns in log parsing, as log messages often contain sensitive information such as IP addresses and domain names~\citep{aghili2025protecting,ekhlasi2025insightai,sallou2024breaking}. 

The significant time consumption and potential privacy issues of semantic-based log parsers pose a problem in their implementations to real-world scenarios, where these modern software systems create large volumes of log data~\citep{yao2021improving, wang2022spine}. With this consideration, researchers made efforts to improve the statistic-based parser's parsing performance by explicitly incorporating generalizable knowledge into the preprocessing or postprocessing stage. \citet{qin2025preprocessing} discovered that most of the variables could be identified in the preprocessing stage, and proposed a general preprocessing framework to enhance parsing performances. \citet{fu2022investigating} merged similar templates to reduce template redundancy. Although these refinement techniques are proven to be effective for statistic-based parsers, we noticed that there is still an effectiveness gap between the two types of parsers. 

Moreover, \citet{chu2021prefix} observed that most statistic-based log parsers rely on data-dependent configurations, making it difficult for users to choose optimal parameters without prior dataset knowledge. \citet{dai2023pilar} further showed that small configuration changes to these statistic-based parsers can lead to significantly different parsing results on the same dataset, and that optimal settings vary across datasets.

To solve these issues, we propose a new unsupervised statistic-based log parser, PIPLUP, in this research. PIPLUP (\textbf{P}lug \textbf{I}t and \textbf{P}lay on \textbf{L}ogs: config\textbf{U}ration-free \textbf{P}arser) parses logs based on a set of data-insensitive parameters, which shows high performance and good generalizability in our experiments
. 
In contrast to existing log parsers that rely on dataset-specific parameters, PIPLUP leverages message-specific 
parameters and globally-insensitive parameters which eliminate the need for configurations and enable the ``plug and play'' feature on new log files, a dynamic constant token searching strategy to effectively group the logs, and an enhanced template extraction and merging process. Our experiment on 14 different datasets and 7 state-of-the-art log parsers illustrates that PIPLUP is significantly more efficient than the state-of-the-art statistic-based parsers and can achieve statistically optimal performance on grouping accuracy (GA) or near-optimal performance on parsing accuracy (PA), F1-score of grouping accuracy (FGA), and F1-score of template accuracy (FTA), even compared to the four state-of-the-art unsupervised semantic-based parsers (i.e., LILAC, LibreLog, LogBatcher, and LUNAR). We aim to comprehensively evaluate the capability of PIPLUP by answering the following research questions:

\noindent\textbf{RQ1: How do different parameters impact PIPLUP?} 
\\
PIPLUP aims to provide a set of data-insensitive configurations for easy deployment. We used a subset of Loghub 2.0~\citep{jiang2024large} to empirically evaluate the impact of the parameters and select the default configuration. 
We observe that our parameters are either data-insensitive or can be dynamically determined by message-level characteristics.

\noindent\textbf{RQ2: How does PIPLUP compare to state-of-the-art parsers in terms of parsing effectiveness?}
\\
We use four standard metrics (GA, FGA, PA, and FTA) to evaluate the effectiveness of PIPLUP and compare it with the baselines. 
We observe that PIPLUP is highly accurate and robust, achieving an overall better performance than state-of-the-art statistic-based parsers and has a competitive performance in comparison to the most effective semantic-based parser (i.e., LUNAR).


\noindent\textbf{RQ3: How does PIPLUP compare to state-of-the-art parsers in terms of parsing efficiency?}
\\
Efficiency is critical for the practical usability of log parsers when processing large files. 
Our evaluation on the Loghub 2.0 benchmark indicates that PIPLUP is efficient at log parsing, with time consumption comparable to state-of-the-art statistic-based parsers and much faster than semantic-based ones.

Our work makes the following main contributions: 
\begin{enumerate} [leftmargin=*]

    \item We proposed a configuration-free log parser, PIPLUP. PIPLUP leverages a novel tree structure without the pre-assumption of the position of constant tokens, and enhances the template extraction approach based on template similarity and describability. Further, it uses a set of data-insensitive parameters, which enables users to directly ``plug and play'' on their log files without excessive configuration needs. 
     \item We systematically evaluated the performance of PIPLUP using the popular Loghub 2.0 benchmark with 14 log files of different systems with varying sizes, demonstrating the high effectiveness and efficiency of our approach. 
    \item We examined the data-insensitivity of PIPLUP's parameters with a subset of Loghub-2.0 and validated the generalizability of the default settings with the full dataset. 
\end{enumerate}

The data and source code of PIPLUP are published in a replication package\footnote{\url{https://github.com/mooselab/PIPLUP-A-Configuration-Free-Statistic-Based-Log-Parser}} for future replication or extension.

The remainder of our paper is organized as follows. 
Section~\ref{sec:motivation} introduces the background of our study. Section~\ref{sec:methodology} describes the structure of PIPLUP and the detailed steps for log parsing. 
Section~\ref{sec:experiment_design} introduces our experiment design and implementation details. Section~\ref{sec:experiment_results} answers our three research questions with the experiment results. Section ~\ref{sec:discussions} provides suggestions for practitioners and researchers on selecting suitable log parsers based on their features. Section~\ref{sec:validity_threat} discusses the threats to the validity of our study and our mitigation strategies, and Section~\ref{sec:related_works} discusses previous studies related to our work. Finally, Section~\ref{sec:conclusion_future_work} concludes our work. 

\section{Background}
\label{sec:motivation}


\subsection{Log Parsing}
\label{sec:log_parsing_importance}

Software systems generate vast amounts of log data~\citep{yao2021improving, wang2022spine}, which provide valuable runtime insights for monitoring system status~\citep{li2020qualitative, liao2020using, chow2014mystery}, ensuring integrity~\citep{lu2017log, yuan2012characterizing, fu2013contextual, messaoudi2021log, nagaraj2012structured, kuttal2011history, gadler2017mining}, and detecting anomalies~\citep{kuttal2011history, gadler2017mining, fu2009execution, he2016experience, du2017deeplog, chen2021experience, wu2023effectiveness}.  However, manually analyzing raw log lines is impractical due to their volume. Log parsing simplifies this process by distinguishing between dynamic variables and static content and extracting structured event templates. This involves two steps: \textbf{log clustering}, where similar messages are clustered, ideally by template, and \textbf{template extraction}, where variables are masked while retaining constant tokens. For example, a log message ``creating device node /udev/vcs7’’ is parsed into the template ``creating device node \textless*\textgreater’’, with the variable ``/udev/vcs7’’ replaced by a predefined placeholder ``\textless*\textgreater’’. These templates can be used for downstream tasks. 

\subsection{Trade-off Between Parsing Effectiveness, Efficiency, and Privacy Concerns}


\label{sec:assumption_defects}

Log parsers are evaluated based on effectiveness and efficiency. \citet{jiang2024large} introduced a large dataset, Loghub 2.0, to assess 15 state-of-the-art log parsers, revealing that only 6 could process all datasets within a practical 12-hour limit. Among them, the statistic-based parsers exhibited faster performance but were outperformed in accuracy by semantic-based parsers like UniParser~\citep{liu2022uniparser} and LogPPT~\citep{le2023log} due to limitations in their heuristics. For example, Drain~\citep{he2017drain} assumes that log messages of the same length and $n$ prefixes belong to the same event cluster. Although it is usually effective, this assumption does not always hold in practice: in the Loghub 2.0 dataset, 11 templates in \textit{Hadoop} start with a variable, and 43.575\% log messages in \textit{OpenStack} share a template with a variable as the first token. Recent semantic-based parsers, such as LILAC~\citep{jiang2024lilac} and LibreLog~\citep{ma2024librelog}, have improved parsing effectiveness by leveraging large language models (LLMs), but they require significantly more time, even with GPU acceleration: when running offline, LILAC requires up to 4 hours to parse ThunderBird, and LibreLog needs around 2.5 hours, approximately $7.9\times$ of Drain's~\citep{ma2024librelog} consumption. Additionally, a recent study by \citet{aghili2025protecting} revealed that system logs contain a high volume of sensitive information, such as IP addresses and host names. Log parsing with closed-source LLMs such as ChatGPT raises privacy concerns, which limits their industrial applicability~\citep{sallou2024breaking, ekhlasi2025insightai}. 

\subsection{Configuration Impacts on Statistic-based Log Parsers}
\label{sec:log_parser_configurations}

The performance of statistic-based log parsers relies heavily on their parameter configurations~\citep{chu2021prefix,dai2023pilar}. A study by \citet{dai2023pilar} illustrated that most of the statistic-based log parsers would provide highly different parsing results on the same dataset when their configurations vary: \textbf{the best performing parser can also become the least effective on the same dataset through changing the parameters}. Based on the findings, Dai \textit{et al.} promoted using generalizable, data-insensitive parameters for better actionability. We categorize the parameters commonly involved in existing methods into three types as follows:

\begin{enumerate} [leftmargin=*]
    \item \textbf{Message-specific parameters.} Message-specific parameters are the ones that are configured on the intrinsic characteristics of individual log messages. For instance, we shall not expect a log message with two tokens to have two variables, as it will degenerate to ``no specific log events''. According to a preliminary analysis of log data, the number of variables in logs is correlated to log lengths. Hence, it is difficult to select a fixed threshold for all logs. Instead, log similarity thresholds can be dynamically determined by their correlation to log lengths. 
    \item \textbf{Dataset-specific parameters.} Dataset-specific parameters are configured differently according to the log dataset being parsed. Take the prefix length $n$ in Drain as an instance, $n$ for OpenSSH is set to 3, but is set to 2 for Apache, because they are optimal based on experiments. However, according to the evaluation conducted by \citet{dai2023pilar}, these parameters would provide highly different parsing results on the same dataset when their configurations vary, suggesting that it is highly challenging to find a unified default setting for them. Further, ground truths for pilot experiments in parameter tuning are usually unavailable in real-life scenarios, posing challenges for users in configuring parses to parse their own log files.
    \item \textbf{Globally-insensitive parameters.} 
    Globally-insensitive parameters are data-insensitive, providing stable and optimal or near-optimal outcomes across different datasets. The entropy threshold leveraged by PILAR~\citep{dai2023pilar} is an example of globally-insensitive parameters: the constant parameter value can achieve (near-)optimal performance across different log datasets. 
\end{enumerate}

According to the definition, both message-specific and globally-insensitive parameters are generalizable and data-insensitive. Using these two types of parameters could protect the actionability and stability of log parsers, making them configuration-free when parsing log files from any system.


\section{Methodology}
\label{sec:methodology}

PIPLUP uses a tree structure to cluster log messages. When a new log message arrives, PIPLUP searches through the dictionary to identify the right cluster for the incoming log message. The dictionary is continuously updated to support online parsing. Below, we first describe the tree structure in Sec.~\ref{sec:tree_structure}, and then we describe the details of our clustering in Sec.~\ref{sec:cluster_searching} and cluster updating strategies in Sec.~\ref{sec:cluster_updating}. Finally, we briefly discuss the template matching process for result output in Sec.~\ref{sec:template_matching}.

\begin{figure*}[h!]
  \center
  \includegraphics[width=\textwidth]{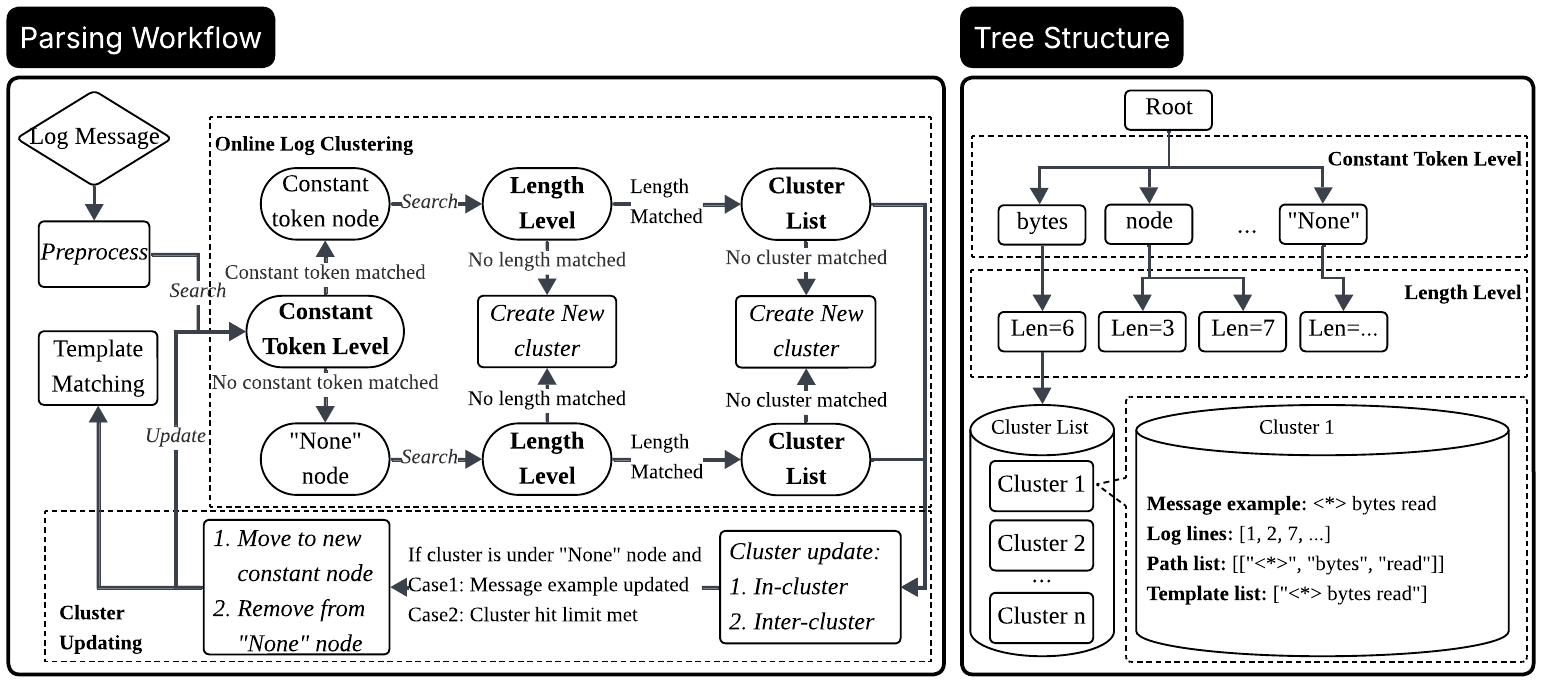}
  \vspace{-2mm}
  \caption{Structure overview of PIPLUP.}
  \label{fig:overview}
  \vspace{-3mm}
\end{figure*}


\subsection{Tree Structure for Log Cluster Representation}
\label{sec:tree_structure}
Inspired by Drain and other tree-based log parsing approaches~\citep{he2017drain,liu2024xdrain,fu2022investigating,qin2025preprocessing,ma2024librelog}, we use a tree structure to represent the clustering of log messages. Our tree structure is illustrated in Fig.~\ref{fig:overview}. Different from Drain, which leverages $n$-prefix for cluster separation, our approach uses a dynamically identified constant token for log clustering. The design aims to overcome the challenge of wrongly clustering event templates that start with a variable and eliminate the need for the predefined, dataset-specific parameter $n$. As shown in Fig.~\ref{fig:overview}, we use the \textbf{Constant Token Level} and \textbf{Length Level} to preliminarily distinguish log messages. The structure is implemented as a two-level dictionary: the first-level keys are constant tokens extracted from the log messages; the second-level keys are the number of tokens of each tokenized log message associated with that constant token; the second-level values are lists of clusters containing the log messages that share the same constant token and length. For example, log messages ``0 bytes read'' and ``128 bytes read'' will be in the same branches since they have the same constant token (i.e., bytes) and length (i.e., 3), while ``node 2: open'' and ``node 2: tried to open, but failed'' will be in different branches since their length is different. 

The log messages under the same branch (same constant tokens and length) are then further split into different clusters based on a similarity threshold $\theta_{sim}$: the log messages with a similarity larger than the threshold are grouped into one cluster. Each cluster is represented by four components: a message example, a log line list, a path list, and a template list. The message example is used for similarity comparison during the searching stage, and will be updated if an incoming message has different tokens from it at the same position: the position will be regarded as having potential variables, thus the token in the example will be replaced with a placeholder. The indices of the logs in a cluster are stored in the log line list. The path list and template list document the token combinations (i.e., different paths) and event templates, summarized with the cluster's information. 

Our design allows multiple templates to be in a single cluster, in order to preserve semantically different events with similar combinations of tokens. Ideally, a cluster of logs should share the same template~\citep{jiang2008abstracting,jiang2024lilac,he2017drain}. However, log clusters with messages that are short in length may contain multiple similar events. Consider setting $\theta_{sim}$ to 0.5, although short messages with variables can be merged together (e.g., ``node: 0 started'', ``node: 1 started'', and ``node: 2 started''), it will also merge highly similar short messages without variables (e.g., ``Opening file, done'' and ``Closing file, done'': the two messages capture different events ``opening'' and ``closing'' in the system). 

This example motivates the importance of preserving templates with similar features but indicating different events. One approach is using semantic analysis with knowledge bases or dictionaries: tokens identified as verbs or statuses should be protected because they tend to document different system behaviors. However, many coding-specific terms (e.g., shutdown, HTTPS, SOCKS5) are not found in standardized dictionaries, making it difficult to apply effective semantic analysis using this method.

Intuitively, if a token contains dynamic variables, the contents from different log messages at the corresponding position should be diverse, and lead to many token sequence combinations in the cluster~\citep{dai2020logram,vaarandi2015logcluster,vaarandi2003data,nagappan2010abstracting}. Hence, we set a globally-insensitive branching threshold $\theta_{br}$ for the number of unique templates to allow the cluster to separate wrongly merged different events. 

When a new log message is input to PIPUP, it is processed with three stages: \textbf{online log clustering}, \textbf{cluster updating}, and \textbf{template matching}. During the log clustering stage, PIPLUP searches for the most compatible cluster for the incoming message using the key on the two dictionary levels and log similarity; if no cluster is found, a new cluster will be created. Then, the cluster's message example and template list will be updated using the message. After updating the cluster, a suitable template will be assigned to the log message during the template matching stage.

\subsection{Online Log Clustering}
\label{sec:cluster_searching}



\subsubsection{\textbf{\underline{Log Preprocessing}}}
\label{sec:tree_step1}
Log preprocessing refers to the process of detecting and masking variables (e.g., replacing the variable tokens with a wildcard ``$\textless$*$\textgreater$'') using domain-knowledge regexes. It has been widely practiced in statistic-based log parsers, as it is known to have the ability to elevate parsing performance~\citep{he2016evaluation,he2017drain,qin2025preprocessing}. Further, a recent study by Qin \textit{et al.} showed that most variables are system generalizable, such as IPv4 addresses and URLs~\citep{qin2025preprocessing}, and can be directly identified during preprocessing. Based on this discovery, Qin \textit{et al.} proposed a general framework for preprocessing, which can enhance the performance of different log parsers using the default regex sets. Hence, we implemented their framework with original settings in PIPLUP to preprocess log messages, and replace the identified variable tokens with ``$\textless$*$\textgreater$''. After preprocessing, the log messages are tokenized with blank spaces. 

\subsubsection{\textbf{\underline{Matching by Constant Tokens}}}
\label{sec:tree_step3}

As mentioned in Sec.~\ref{sec:assumption_defects}, although it is not a common case among all datasets, some events have a variable as their first token. Therefore, using n-prefixes (i.e., the first n tokens) as a clustering indicator may lead to unstable clustering results; further, it requires adequate domain knowledge to fix the length of the static prefix string. Hence, we propose to dynamically update the constant token keys (i.e., the dictionary keys on the constant token level) and create coarse clusters of logs. 

When a new log message is input to PIPLUP, the system searches through the list of constant tokens (i.e., the first-level keys, as mentioned in Sec.~\ref{sec:tree_structure}) to find a match within the message. Specifically, it checks each constant token one by one to see if that token appears anywhere in the log message. Once a matching constant token is found, it is assigned as the message’s constant token, and the search stops immediately. If no constant token from the list is found in the message, the message is associated with a special "None" node, indicating that no constant token can represent the message at this layer. In this case, the cluster remains pending further constant token searching.



The constant token of a cluster under the ``None'' node will be identified in two cases: \textbf{message example is updated} and \textbf{hit limit is met}. In the first case (i.e., example template is updated), as introduced in Sec.~\ref{sec:tree_structure}, a change in the message example indicates some potential variables are detected.
According to the previous study by \citet{qin2025preprocessing}, only a small portion of variables cannot be identified during the preprocessing stage. Therefore, we assume that potential variables are eliminated (i.e., replaced by placeholders) if the example is updated. 
In the second case (i.e., the hit limit is met), we set a default hit limit based on the sample size needed for a 95\% confidence level and 5\% error margin (i.e., 385 for unlimited population size). If adequate new log messages are appended to this cluster without triggering a message example update, 
we assume that the existing message example does not contain extra variables by 95\% confidence level. In both cases, the first token in the message example without a placeholder will be identified as a constant: the selection of the token is based on the commonly effective assumption from Drain, that ``tokens in the beginning positions are more likely to be constants''~\citep{he2017drain}. 

Once a new constant token is identified, PIPLUP creates a new constant node and performs a cluster searching process under the ``None'' node. If a cluster's message example contains the new constant, the cluster will be removed from the ``None'' node and appended to the new constant node. 

\subsubsection{\textbf{\underline{Matching by Length}}}
\label{sec:tree_step4}
Although constant tokens can be efficient in describing templates, they are not adequate for constructing fine-grained clusters. For instance, ``node: 1 started'' and ``node: 1 started, closing in 1 sec'' share the same constant token ``node:'', but they do not belong to the same log cluster. To split the clusters into finer granularity, we further divide the clusters according to their lengths (i.e., the number of tokens) on the \textbf{Length Layer}. Based on the intuition of \citet{he2017drain}, log messages sharing the same template tend to have the same length. We acknowledge that there exist log messages with different lengths belonging to the same event (e.g., both ``opening Chrome.exe'' and ``opening Microsoft Word.exe'' match with ``opening \textless*\textgreater''); the elimination of these issues is attempted during template merging described in Sec.~\ref{sec:inter_cluster_updates}. 

\subsubsection{\textbf{\underline{Matching by Log Similarity}}}
\label{sec:tree_step5}
Following the clustering procedure in the previous steps, a log message shall reach the leaf node of PIPLUP's parsing tree in the current step. Through message similarity clustering, PIPLUP calculates the similarity between the log message and each cluster's message example ($Ex_c$). Similar to Drain, log similarity is defined as the proportion of common tokens between the two log messages. The log message shall be assigned to the log cluster with the largest similarity on the leaf node if the similarity value is larger than a predefined threshold. The formula for $Sim(log, example)$ writes as:

\[
Sim(log, Ex_c)= \frac{\sum_{i=1}^{l} Equ(log(i), Ex_c(i))}{l}
\]
\noindent where 
\[
Equ(t_1, t_2)=
\begin{cases}
    1, & t_1 == t_2\\
    0, & otherwise
\end{cases}
\]


\noindent $l$ indicates the length of the two token sequences. $log(i)$ and $Ex_c(i)$ represent the $i^{th}$ token of the compared log message and the message example, respectively. Given that a length-based search is done in the last step, the log message and cluster token list shall have the same length. $Equ(t_1, t_2)$ is an indicator function to determine whether two tokens $t_1$ and $t_2$ are the same. The formula $Sim(seq_1, seq_2)$ measures the similarity between two log sequences by calculating the portion of shared tokens (i.e., $seq_1(i)$ and $seq_2(i)$). After calculating the similarities between the log message and all log clusters' message examples on the leaf, we extract the log cluster with the largest $Sim(seq_1, seq_2)$ value. The log message will be regarded as belonging to the cluster if the value is larger than a predefined threshold $\theta_{sim}$ (i.e., $Sim(seq_1, seq_2)>\theta_{sim}$).

$\theta_{sim}$ is used to control the granularity of log merging. A smaller $\theta_{sim}$ allows two log messages with larger differences belonging to one cluster, and vice versa. When $\theta_{sim}$ is set to 0, log messages with the same constant tokens and length layers are merged together. In contrast, if $\theta_{sim}$ is increased to 1, the log clusters no longer tolerate additional variables apart from preprocessed ones, and the leaves will contain numerous tiny clusters
. Neither extremely loose nor strict granularity is suitable for log parsing. Therefore, the value of $\theta_{sim}$ shall be predefined with care. To eliminate the need for extra efforts, As previously discussed in Sec.~\ref{sec:log_parser_configurations}, the number of variables is empirically related to log lengths. Hence, PIPLUP uses a set of \textbf{message-specific} parameters that are dynamically adjusted according to log lengths, $\theta_{sim}$, for similarity comparison. The detailed threshold setting heuristics are addressed in Sec.~\ref{sec:rq1}.

If the maximum $Sim(seq_1, seq_2)$ is larger than $\theta_{sim}$, we consider the log message to belong to the cluster corresponding to the highest similarity. Otherwise, the log is not matched with any existing cluster. In this case, PIPLUP creates a new cluster based on the message and stores it under the same leaf of the same length, and we say that the message is matched with an empty cluster.




\subsection{Cluster Updating}
\label{sec:cluster_updating}
Following the cluster searching process in Sec.~\ref{sec:cluster_searching}, a cluster is matched with the input log message. The matched cluster is dynamically updated to allow online parsing. First, an in-cluster update will be triggered to append to the log line number the cluster's log line list and update the cluster's message example, path list, and template list. The path list records different token combinations (i.e., paths) in existing log lines. Ultimately, the path list is used to update the template list of the cluster. If the template list is updated, an inter-cluster template merging process among clusters under the same constant token node will be triggered to reduce redundancy in event templates. 


\subsubsection{\textbf{\underline{
In-cluster Updates}}}
\label{sec:in_cluster_updates}

As mentioned in Sec.~\ref{sec:tree_step5}, the message example is used to measure the similarity between the incoming log message and log clusters on its corresponding leaf node. For log messages matching an empty cluster (i.e., no existing cluster is matched in the last stage), the preprocessed log message will be used as the template of the cluster. If an existing cluster is matched, PIPLUP compares the tokens in the log message and the cluster's message example. The token in the example will be changed to ``\textless*\textgreater'' if it doesn't match the token in the log message at the same position.



Then, PIPLUP scans the log message's token sequence and examines its describability by the existing token paths. The describability is defined as follows: at each position, each token in the sequence matches the corresponding token in the path, or the corresponding token in the path is already identified as a variable. If the message cannot be described by any existing path, PIPLUP will append the new log message into the path list if the number of paths is smaller than the template branching threshold $\theta_{br}$. Otherwise, PIPLUP will search for the most similar path using the formula defined in Sec.~\ref{sec:tree_step5} for path update. The most similar path will be compared with the incoming sequence: if the tokens at a specific position are different, PIPLUP identifies it as a variable. Once a variable is identified, the tokens in this position for all the templates will be replaced with a placeholder. Take the same example in Sec.~\ref{sec:tree_structure} as an instance, when $\theta_{br}$
is set to 2, the cluster containing messages ``node: 0 started'', ``node: 1 started'', and ``node: 2 started'' will have one template ``node: \textless*\textgreater started'', and the other cluster containing ``Opening file, done'' and ``Closing file, done'' will retain two different templates. The evaluation of the default $\theta_{br}$ is discussed in Sec.~\ref{sec:rq1}.

After updating the path lists, we join the tokens in each path with blank spaces to obtain new templates. Jiang \textit{et al.} stressed the importance of readability and conciseness of event templates in LogHub 2.0~\citep{jiang2024large}. The researchers agreed on the heuristic that if two templates contain the same constant tokens and different numbers of variables, the one with fewer ``\textless*\textgreater'' should be chosen. Following this guidance, we correct the generated templates by merging repetitive wildcards. We leveraged the template correction rules provided by LOGPAI~\citep{huang2025no} to postprocess the templates and update the cluster's template list.



\subsubsection{\textbf{\underline{Inter-Cluster Updates}}}
\label{sec:inter_cluster_updates}


Fu \textit{et al.} discovered that statistic-based log parsers tend to have unsatisfactory performance when deployed in industrial scenarios~\citep{fu2022investigating}. According to their study, an event merging process is needed to merge log messages of different lengths sharing the same template. The merging process based on Jaccard similarity is shown to increase the average grouping accuracy across 16 datasets with a globally-insensitive threshold of 0.6: if the Jaccard similarity between two templates are higher than 0.6, then they are merged by replacing the different tokens with a wildcard. However, testing mergeability with the approach proposed by \citet{fu2022investigating} may lead to wrong merges when the event templates have different token sequences. For instance, ``Done closing file'' and ``Closing file done'' are regarded as highly similar according to this criterion.

We thus take an alternative approach by testing the describability between two event templates. First, we calculate the Jaccard similarity between two log templates; if their similarity is higher than the threshold (i.e., 0.6, as justified by Fu \textit{et al.}), then we further examine their describability. If the template of event A can also be matched with the template of event B, we say event A can be described by event B, and event A can be merged with event B. The merging process is only done among clusters with the same constant token but different lengths. Take the example previously addressed in Sec.~\ref{sec:tree_step4} for instance: assume ``opening Chrome.exe'' has the template ``opening \textless*\textgreater'', but ``opening Microsoft Word.exe'' remained the same after template inference. Since the template ``opening \textless*\textgreater'' can describe the latter log template, a false positive template ``opening Microsoft Word.exe'' will be replaced by the more concise template. The inter-cluster merging function is triggered only when a cluster's template list is updated. 


\subsection{Template Matching}
\label{sec:template_matching}
Following the procedure in Sec.~\ref{sec:in_cluster_updates}, PIPLUP may create multiple templates for a log cluster. If a cluster contains only one template, then the log message will be matched with this event directly. Conversely, if multiple templates are inferred, we assign a suitable template to the message using regex matching (regex derived from templates). 

\section{Experiment Design}
\label{sec:experiment_design}

\subsection{Dataset}
\label{sec:dataset}
We use Loghub 2.0~\citep{jiang2024large} in the experiments: the number of annotated log lines in Loghub 2.0 ranges from 23,921 to 16,601,745, and the number of templates for each system varies from 11 to 1,241. Before starting the experiment, we corrected the event templates with the latest rules provided by LOGPAI~\citep{huang2025no} to ensure the quality of the ground truths. The dataset is widely used in recent log parsing studies~\citep{ma2024librelog,jiang2024lilac,qin2025preprocessing}. We use a subset of log files in Loghub 2.0 to understand the impact of the parameters ($\theta_{sim}$ and $\theta_{br}$) on PIPLUP's parsing performance. 


\subsection{Baseline Parser Selection}
\label{sec:parser_selection}

To understand the effectiveness and efficiency of PIPLUP, we compare its performance with seven state-of-the-art unsupervised log parsers, including both statistic-based and semantic-based ones. We exclude supervised log parsers (e.g., few-shot LILAC~\citep{jiang2024lilac} and UNLEASH~\citep{le2025unleashing}) as our baselines since they require domain experts' manual labeling for each individual dataset, which is undesirable for fully automated parsing. Below, we briefly introduce our baselines.

\noindent\underline{\textbf{Drain}}~\citep{he2017drain}. According to the large-scale evaluation by \citet{jiang2024large}, Drain is on average the most effective and one of the most efficient statistic-based parsers. We use the implementation of Drain provided along with Loghub 2.0, with the default parameter settings for each dataset. 

\noindent\underline{\textbf{XDrain}}~\citep{liu2024xdrain}. XDrain inherits the fixed-depth tree design of Drain, but leverages random shuffling of tokens to mitigate parsing failures caused by logs starting with variables. Further, it generates a parsing forest for each log to select the most robust template by voting. We used the replication code~\citep{liu2024xdrain} with default settings and evaluated the performance with the Loghub 2.0 benchmark.

\noindent\underline{\textbf{Preprocessed-Drain}}~\citep{qin2025preprocessing}. A recent study by \citet{qin2025preprocessing} improves statistic-based parsers such as Drain's effectiveness with a lightweight log preprocessing step. We use their replication code with default settings to obtain the evaluation results. 
    
\noindent\underline{\textbf{LILAC}}~\citep{jiang2024lilac}. LILAC leverages ChatGPT (GPT-3.5-Turbo-0613) for effective log parsing. Through a set of experiments, \citet{jiang2024lilac} illustrated that ChatGPT is highly effective in log parsing. 
Given that GPT-3.5-Turbo-0613 is no longer available, we leveraged GPT-3.5-Turbo-0125 while replicating the 0-shot results. 

\noindent\underline{\textbf{LibreLog}}~\citep{ma2024librelog}. LibreLog uses smaller-scaled open-source LLMs (e.g., Llama3-8B, Mistral-7B) for log parsing. The unsupervised semantic-based log parser uses a template memory to achieve better parsing efficiency than LILAC. 
Due to resource limitations, we obtained the parsing results of LibreLog when paired with Llama3-8B (i.e., the most capable setting reported in the original study) from the authors and evaluated the performance using the corrected ground truths. 

\noindent\underline{\textbf{LogBatcher}}~\citep{xiao2024free}. LogBatcher contains three components for log parsing: partitioning, caching, and querying, which allow the parser to effectively and efficiently parse logs. We leveraged the code with the default settings (i.e., using GPT-3.5-Turbo-0125) documented in the paper for replication. 

\noindent\underline{\textbf{LUNAR}}~\citep{huang2025no}. LUNAR leverages the commonality and variability among logs and clusters them into Log Contrastive Units (LCUs). According to a set of experiments, LUNAR significantly outperforms state-of-the-art log parsers. We replicated the results with the code's default settings and called GPT-3.5-turbo-0125 for log parsing.

We did not implement Drain+~\citep{fu2022investigating} and Prefix-Graph~\citep{chu2021prefix} in this study, as their source code are not publicly available. Although PILAR~\citep{dai2023pilar} achieved the most stable performance across datasets, its average accuracy failed to exceed Drain under the default settings. Hence, it is not compared in our experiment; detailed PILAR evaluation results can be found in our replication package. 

\subsection{Evaluation Metrics}
\label{sec:evaluation_metrics}

Two representative metrics, grouping accuracy (GA) and parsing accuracy (PA), are commonly used for log parsing evaluations at the log message level~\citep{zhu2019tools, dai2020logram,khan2022guidelines,he2017drain,ma2024librelog}. \citet{khan2022guidelines} further extend PA on log template level and established F1-score of template accuracy (FTA), and Jiang \textit{et al.} evaluated the template-level grouping performance with F1-score of group accuracy (FGA)~\citep{jiang2024large}. While GA and PA are widely used in previous studies, FGA and FTA are less sensitive to imbalanced template frequencies and can offer a more robust evaluation result, as rare log events, which may not significantly impact PA or GA, may carry important information (e.g., anomalies). Hence, we combine the four metrics to understand PIPLUP's performance comprehensively. 

\noindent\underline{\textbf{Grouping Accuracy (GA)}}~\citep{zhu2019tools}. A ``correct group'' is defined as log messages with the same ground truth template clustered into one group. Based on this definition, GA is further defined as the ratio of correctly grouped log messages in the dataset by the log parser. It evaluates the parser's ability to group log messages on the message level. 

\noindent\underline{\textbf{Parsing Accuracy (PA)}}~\citep{dai2020logram}. A ``correct parse'' suggests that all the variables in a log message are spotted, none of the constant tokens are misidentified as variables, and the summarized template is identical to the ground truth. PA calculates the portion of correctly parsed messages. It evaluates the parser's ability to detect the variables on the message level. 

\noindent\underline{\textbf{F1-score of Group Accuracy (FGA)}}~\citep{jiang2024large}. A ``correctly grouped template'' is defined as all log messages sharing the same ground truth template clustered into one group. FGA considers the portion of correctly grouped templates in the dataset and is the harmonic mean of the precision (PGA) and recall of group accuracy (RGA). PGA and RGA are calculated as 
$PGA=\frac{N_c}{N_p},\ RGA=\frac{N_c}{N_g}$, 
where $N_c$ represents the number of templates correctly grouped by the parser, $N_p$ represents the number of groups identified by the parser, and $N_g$ stands for the number of groups in the ground truth. Based on the two metrics, FGA is defined as $FGA=\frac{2*PGA*RGA}{PGA+RGA}$. It evaluates the parser's ability to group log messages on the template level. 

\noindent\underline{\textbf{F1-score of Template Accuracy (FTA)}}~\citep{khan2022guidelines}. A ``correctly identified template'' should satisfy two criteria: 1) the inferred template is identical to the ground truth; 2) the template should be correctly grouped, as defined in FGA. FTA represents the F1-score of correctly identified templates by the parser. It is the harmonic mean of the precision (PTA) and recall of template accuracy (RTA). Similar to FGA, we define the two metrics with $PTA=\frac{\hat{N_c}}{N_p},\ RTA=\frac{\hat{N_c}}{N_g}$, where $\hat{N_c}$ is the number of correctly identified templates. FTA shall then be written as $FTA=\frac{2*PTA*RTA}{PTA+RTA}$. It evaluates the parser's ability to detect the variables on the template level. 

\noindent\textbf{Comparing log parsers}. We compare PIPLUP with the baselines based on these metrics, as well as their \textbf{average} and \textbf{variance} across the log datasets. We also use the \textbf{Scott-Knott effect size difference (ESD)}~\citep{tantithamthavorn2018impact} to statistically rank the log parsers based on the distribution of the metric values across the log datasets: log parsers exhibiting no statistically significant difference (p-value$>$0.05) or with \emph{negligible} difference in terms of a metric are ranked in the same place. 

\subsection{Experiment Environment}
\label{sec:implementation_environment}
The experiments are conducted on a Mac Mini with an M2 chip and 16GB of memory. 
Apart from PIPLUP, we also replicate the results of all the log parsers except LibreLog. 
Due to resource limitations, we did not replicate LibreLog in this study, as our evaluation datasets are the same as the ones used in the original papers of  LibreLog~\citep{ma2024librelog}. The tool uses an Ubuntu server with NVIDIA Tesla A100 GPU, AMD EPYC 7763 64-core CPU, and 256GB RAM. 
\section{Experiment Results}
\label{sec:experiment_results}

\subsection*{\textbf{RQ1: How different parameters impact PIPLUP?}}
\label{sec:rq1}

As introduced in Sec~\ref{sec:methodology}, two parameters are required in PIPLUP's parsing process, namely $\theta_{sim}$ and $\theta_{br}$. $\theta_{sim}$ is a message-specific parameter used for similarity-based clustering (Sec.~\ref{sec:tree_step5}); $\theta_{br}$ is a globally-insensitive parameter used as a branching threshold to preserve template variety (Sec.~\ref{sec:tree_structure}). Based on empirical experiments on the 10 smallest log files from Loghub 2.0, we determined the default settings for PIPLUP: $\theta_{sim}=dynamic$ and $\theta_{br}=2$. To evaluate the effectiveness of these settings, we compare them with other parameter candidates using the same 10 smallest log files. 
\textbf{When we vary one parameter, we fix the other parameter at its default value.}
The remaining files (i.e., Spark, Thunderbird, BGL, and HDFS) are used to test generalizability in RQ2 and RQ3. 


\noindent\underline{\textbf{Evaluating the impact of $\theta_{sim}$}}. As mentioned in Sec.~\ref{sec:tree_step5}, the similarity threshold $\theta_{sim}$ of PIPLUP is dynamically adjusted based on the length of individual log messages rather than the dataset. Given that the log lengths are not constrained, we collect the log template lengths in the dataset and split them into four groups according to quartiles: the first quartile (25\%) consists of short messages with no more than 4 tokens; the second quartile (50\%) included mid-length messages which contains 5 or 6 tokens; the third quartile (75\%) involves longer messages with 7 to 9 tokens; the last quartile has messages longer than 9 tokens. We then, empirically, used the third quartile of the variable numbers in the first three groups, and set the representative number of variables to 1, 2, and 3, respectively. For messages with only 1 token, we change the allowance to 0 to avoid wrongly merging all messages of the same length of one token. $\theta_{sim}$ is then calculated based on the log length and variable number, as shown in Table~\ref{tab:sim_thresholds}.  For messages longer than 9, the distribution of variables has a high variance. Therefore, the $\theta_{sim}$ for log messages longer than 9 is relaxed to 0.5, the similarity threshold mode across datasets used in Drain. The detailed settings are listed in Table~\ref{tab:sim_thresholds}. 

We evaluated PIPLUP's performance using multiple static $\theta_{sim}$ values (i.e., 0.4, 0.5, 0.6, and 0.7) and our message-specific, dynamic $\theta_{sim}$ to understand the impact of dynamic similarity thresholds. According to the results in Table~\ref{tab:rq1}, the change in $\theta_{sim}$ can create a significant impact on the average performance: the gap between the best and worst FGA is 17.335\%, and the gap in FTA is 15.385\%. The average PA, FGA, and FTA of using the dynamic setting is the best among all the settings, and the average GA is near-optimal, with a small gap of 1.299\% to the optimal value; for 3 individual datasets, the dynamic setting achieves the best performance in terms of most of the metrics; the average performance gap between the dynamic setting and the best setting is from 1.065\% (for the PA metric) to 2.175\% (for the GA metric). 
We noticed that some datasets prefer specific static thresholds. For instance, setting $\theta_{sim}$ to 0.7 can significantly boost the parsing performance of PIPLUP on OpenStack, and Linux achieved the optimal parsing result when it is set to 0.5. However, the thresholds may not apply to other datasets, given that using the high $\theta_{sim}$ threshold will fail to parse short log messages with variables, and a constant value of 0.5 could wrongly merge short messages (e.g., a log message containing only four tokens). 
If a log template with 3 tokens contains one variable, the maximum log message similarity in this cluster would be 0.67. In comparison, \textbf{our dynamic message-specific (default) setting of $\theta_{sim}$ achieved the optimal or near-optimal performance compared to using fixed thresholds}. 

\noindent\underline{\textbf{Evaluating the impact of $\theta_{br}$}}. $\theta_{br}$ is the branching threshold for templates. 
To get the candidates for $\theta_{br}$, we analyzed the event templates in the 10 log files by clustering them with the dynamic similarity threshold $\theta_{sim}$. The templates are clustered together if their pairwise similarity is below the threshold. In this case, we say that the branching method is required to separate different events. After clustering the templates, we calculate the cluster sizes on each dataset and check the value of their $99^{th}$ percentile (i.e., covers 99\% of the cases); the resulting value ranges from 1 to 6. Therefore, the candidates for $\theta_{br}$ range from 2 to 6, as $\theta_{br}$ should be greater than 1 to be able to distinguish variable from constant tokens (Sec.~\ref{sec:in_cluster_updates}). 
As reported in Table~\ref{tab:rq1}, \textbf{when set within a reasonable range, $\theta_{br}$ is \textit{globally insensitive}}: the average difference between the best and worst is only 1.425\% for GA, 1.026\% for PA, 3.467\% for FGA, and 2.918\% for FTA. The gap is also small on most individual datasets: among the datasets, only OpenStack exhibited preference for a high branching threshold (i.e., $\theta_{br}$=6), which corresponds to our previous calculation, as its template cluster size at $99^{th}$ percentile is 6. On the other hand, the remaining 9 datasets adapted well to the default setting ($\theta_{br}$=2), suggesting a generalizable solution: it achieved the maximum values on 5 datasets, with an average performance gap less than 1\% from the maximum in terms of all four metrics. Our default setting obtained the best performance on average. 

\begin{table}[t]
\centering
\caption{Default $\theta_{sim}$ settings by log length}
\vspace{-1mm}
\label{tab:sim_thresholds}
\resizebox{0.85\columnwidth}{!}{%
\begin{tabular}{l|l|l}
\hline
\textbf{Length Range ($l$)} & \textbf{Representative Variable Number ($n$)} & \textbf{$\theta_{sim}$} \\ \hline
1   & 0 & 1               \\ \hline
2-4 & 1 & $\frac{l-n}{l}$ \\ \hline
5-6 & 2 & $\frac{l-n}{l}$ \\ \hline
7-9 & 3 & $\frac{l-n}{l}$ \\ \hline
9+  & - & 0.5             \\ \hline
\end{tabular}%
}
\vspace{-3mm}
\end{table}

\setlength{\tabcolsep}{3pt}
\begin{table}[]
\centering
\caption{The default setting performance, max and min value of GA, PA, FGA, and FTA achieved by PIPLUP with different settings of $\theta_{sim}$ and $\theta_{br}$ on the 10 datasets. The maximum obtained by PIPLUP's default settings (i.e., $\theta_{sim}=dynamic$ and $\theta_{br}=2$) are highlighted with bold fonts. }
\label{tab:rq1}
\resizebox{\columnwidth}{!}{%
\begin{tabular}{l|l|cccc|cccc}
\hline
\multirow{2}{*}{\textbf{Dataset}} &
  \multirow{2}{*}{\textbf{Config}} &
  \multicolumn{4}{c|}{$\theta_{sim}$} &
  \multicolumn{4}{c}{$\theta_{br}$} \\ \cline{3-10} 
 &
   &
  \textbf{GA} &
  \textbf{PA} &
  \textbf{FGA} &
  \textbf{FTA} &
  \textbf{GA} &
  \textbf{PA} &
  \textbf{FGA} &
  \textbf{FTA} \\ \hline
\multirow{2}{*}{\textbf{Proxifier}} &
  \textbf{Default} &
  \textbf{1.000} &
  \textbf{0.994} &
  \textbf{1.000} &
  \textbf{0.909} &
  \textbf{1.000} &
  \textbf{0.994} &
  \textbf{1.000} &
  \textbf{0.909} \\
 &
  \textbf{Min/Max} &
  1.000/1.000 &
  0.994/0.994 &
  1.000/1.000 &
  0.909/0.909 &
  1.000/1.000 &
  0.994/0.994 &
  1.000/1.000 &
  0.909/0.909 \\ \hline
\multirow{2}{*}{\textbf{Linux}} &
  \textbf{Default} &
  0.867 &
  0.731 &
  0.921 &
  0.722 &
  \textbf{0.867} &
  0.731 &
  0.921 &
  0.722 \\
 &
  \textbf{Min/Max} &
  0.845/0.988 &
  0.720/0.732 &
  0.859/0.931 &
  0.688/0.728 &
  0.813/0.867 &
  0.711/0.734 &
  0.901/0.927 &
  0.719/0.736 \\ \hline
\multirow{2}{*}{\textbf{Apache}} &
  \textbf{Default} &
  \textbf{1.000} &
  \textbf{0.994} &
  \textbf{1.000} &
  \textbf{0.828} &
  \textbf{1.000} &
  \textbf{0.994} &
  \textbf{1.000} &
  \textbf{0.828} \\
 &
  \textbf{Min/Max} &
  0.997/1.000 &
  0.994/0.994 &
  0.918/1.000 &
  0.787/0.828 &
  0.993/1.000 &
  0.991/0.994 &
  0.844/1.000 &
  0.719/0.828 \\ \hline
\multirow{2}{*}{\textbf{Zookeeper}} &
  \textbf{Default} &
  \textbf{0.970} &
  \textbf{0.825} &
  0.873 &
  0.727 &
  \textbf{0.970} &
  \textbf{0.825} &
  0.873 &
  0.727 \\
 &
  \textbf{Min/Max} &
  0.967/0.970 &
  0.822/0.825 &
  0.825/0.914 &
  0.700/0.774 &
  0.963/0.970 &
  0.818/0.825 &
  0.836/0.882 &
  0.701/0.741 \\ \hline
\multirow{2}{*}{\textbf{MAC}} &
  \textbf{Default} &
  \textbf{0.871} &
  \textbf{0.553} &
  0.883 &
  0.551 &
  \textbf{0.871} &
  \textbf{0.553} &
  \textbf{0.883} &
  \textbf{0.551} \\
 &
  \textbf{Min/Max} &
  0.854/0.871 &
  0.548/0.553 &
  0.874/0.887 &
  0.544/0.552 &
  0.859/0.871 &
  0.548/0.553 &
  0.860/0.883 &
  0.536/0.551 \\ \hline
\multirow{2}{*}{\textbf{Hadoop}} &
  \textbf{Default} &
  0.908 &
  \textbf{0.693} &
  0.880 &
  0.574 &
  \textbf{0.908} &
  \textbf{0.693} &
  \textbf{0.880} &
  \textbf{0.574} \\
 &
  \textbf{Min/Max} &
  0.737/0.909 &
  0.522/0.693 &
  0.072/0.894 &
  0.046/0.586 &
  0.862/0.908 &
  0.676/0.693 &
  0.815/0.880 &
  0.537/0.574 \\ \hline
\multirow{2}{*}{\textbf{OpenStack}} &
  \textbf{Default} &
  0.950 &
  0.417 &
  0.946 &
  0.710 &
  0.950 &
  0.417 &
  0.946 &
  0.710 \\
 &
  \textbf{Min/Max} &
  0.950/1.000 &
  0.417/0.466 &
  0.946/1.000 &
  0.710/0.771 &
  0.950/1.000 &
  0.417/0.466 &
  0.946/1.000 &
  0.710/0.771 \\ \hline
\multirow{2}{*}{\textbf{HealthApp}} &
  \textbf{Default} &
  0.996 &
  \textbf{0.971} &
  0.914 &
  \textbf{0.763} &
  0.996 &
  \textbf{0.971} &
  0.914 &
  \textbf{0.763} \\
 &
  \textbf{Min/Max} &
  0.995/0.997 &
  0.971/0.971 &
  0.476/0.926 &
  0.394/0.763 &
  0.996/0.997 &
  0.971/0.971 &
  0.914/0.931 &
  0.755/0.763 \\ \hline
\multirow{2}{*}{\textbf{HPC}} &
  \textbf{Default} &
  \textbf{0.866} &
  \textbf{0.990} &
  \textbf{0.896} &
  \textbf{0.883} &
  \textbf{0.866} &
  \textbf{0.990} &
  \textbf{0.896} &
  \textbf{0.883} \\
 &
  \textbf{Min/Max} &
  0.866/0.866 &
  0.990/0.990 &
  0.896/0.896 &
  0.883/0.883 &
  0.807/0.866 &
  0.933/0.990 &
  0.802/0.896 &
  0.802/0.883 \\ \hline
\multirow{2}{*}{\textbf{OpenSSH}} &
  \textbf{Default} &
  0.691 &
  \textbf{0.633} &
  \textbf{0.921} &
  \textbf{0.868} &
  \textbf{0.691} &
  \textbf{0.633} &
  \textbf{0.921} &
  \textbf{0.868} \\
 &
  \textbf{Min/Max} &
  0.691/0.722 &
  0.443/0.633 &
  0.619/0.921 &
  0.584/0.868 &
  0.691/0.691 &
  0.633/0.633 &
  0.921/0.921 &
  0.868/0.868 \\ \hline
\multirow{2}{*}{\textbf{Average}} &
  \textbf{Default} &
  0.912 &
  \textbf{0.780} &
  \textbf{0.923} &
  \textbf{0.754} &
  \textbf{0.912} &
  \textbf{0.780} &
  \textbf{0.923} &
  \textbf{0.754} \\
 &
  \textbf{Min/Max} &
  0.899/0.924 &
  0.747/0.780 &
  0.763/0.923 &
  0.638/0.754 &
  0.899/0.912 &
  0.772/0.780 &
  0.891/0.923 &
  0.732/0.754 \\ \hline
\end{tabular}%
}
\end{table}

\begin{boxD}
Our impact evaluation of PIPLUP's parameters confirmed our assumptions: our dynamic setting of $\theta_{sim}$ outperforms fixed settings in terms of both performance and robustness; $\theta_{br}$ is globally insensitive across different datasets. Our default settings on the two parameters achieve optimal or near-optimal performance for all datasets.
\end{boxD}

\subsection*{\textbf{RQ2: How does PIPLUP compare to state-of-the-art parsers in terms of parsing effectiveness?}}
\label{sec:rq2}

\begin{table}[]
\centering
\caption{PIPLUP's succeeded and failed template extraction cases under the default branching threshold ($\theta_{br}=2$). } 
\label{tab:branch_cases}
\resizebox{\columnwidth}{!}{%
\begin{tabular}{l|l}
\hline
\textbf{}        & \textbf{Succeeded case}                                                                      \\ \hline
\textbf{branch1} & 
Changing view \textcolor{red}{acls} to: $\textless$*$\textgreater$ \\
\textbf{branch2} & Changing view \textcolor{red}{modify} to: $\textless$*$\textgreater$
\\ \hline
                 & \textbf{Failed case}                                                                         \\ \hline
\textbf{branch1} & Task cleanup failed for attempt \textcolor{red}{attempt\_1445144423722\_0020\_m\_000002\_0}                   \\
\textbf{branch2} & Task cleanup failed for attempt \textcolor{red}{attempt\_1445144423722\_0020\_m\_000001\_0}                   \\ \hline
\end{tabular}%
}
\vspace{-2mm}
\end{table}

\begin{table}[]
\centering
\caption{The effectiveness (GA, PA, FGA, FTA) comparison between the eight log parsers. The best metric results for each type of parsers (i.e., statistic-based and semantic-based) are highlighted with bold fonts, and the worst results are printed in italics. A positive gap with PIPLUP indicates the tool exceeds PIPLUP on the corresponding metric, and vice versa. }
\label{tab:effectiveness}
\resizebox{\columnwidth}{!}{%
\begin{tabular}{l|llllllllllllllll}
\hline
\multirow{3}{*}{\textbf{Dataset}} &
  \multicolumn{16}{c}{\textbf{Statistic-based Parsers}} \\ \cline{2-17} 
 &
  \multicolumn{4}{c|}{\textbf{Drain}} &
  \multicolumn{4}{c|}{\textbf{XDrain}} &
  \multicolumn{4}{c|}{\textbf{Pre-Drain}} &
  \multicolumn{4}{c}{\textbf{PIPLUP}} \\ \cline{2-17} 
 &
  \multicolumn{1}{c}{\textbf{GA}} &
  \multicolumn{1}{c}{\textbf{PA}} &
  \multicolumn{1}{c}{\textbf{FGA}} &
  \multicolumn{1}{c|}{\textbf{FTA}} &
  \multicolumn{1}{c}{\textbf{GA}} &
  \multicolumn{1}{c}{\textbf{PA}} &
  \multicolumn{1}{c}{\textbf{FGA}} &
  \multicolumn{1}{c|}{\textbf{FTA}} &
  \multicolumn{1}{c}{\textbf{GA}} &
  \multicolumn{1}{c}{\textbf{PA}} &
  \multicolumn{1}{c}{\textbf{FGA}} &
  \multicolumn{1}{c|}{\textbf{FTA}} &
  \multicolumn{1}{c}{\textbf{GA}} &
  \multicolumn{1}{c}{\textbf{PA}} &
  \multicolumn{1}{c}{\textbf{FGA}} &
  \multicolumn{1}{c}{\textbf{FTA}} \\ \hline
\textbf{Proxifier} &
  0.692 &
  \textit{0.688} &
  \textit{0.206} &
  \multicolumn{1}{l|}{\textit{0.176}} &
  \textit{0.510} &
  \textit{0.688} &
  0.600 &
  \multicolumn{1}{l|}{0.500} &
  0.989 &
  \textbf{0.994} &
  0.870 &
  \multicolumn{1}{l|}{0.870} &
  \textbf{1.000} &
  \textbf{0.994} &
  \textbf{1.000} &
  \textbf{0.909} \\
\textbf{Linux} &
  \textit{0.686} &
  0.112 &
  \textit{0.778} &
  \multicolumn{1}{l|}{\textit{0.262}} &
  \textbf{0.943} &
  \textit{0.111} &
  0.780 &
  \multicolumn{1}{l|}{0.266} &
  0.710 &
  0.118 &
  0.857 &
  \multicolumn{1}{l|}{0.492} &
  0.867 &
  \textbf{0.731} &
  \textbf{0.921} &
  \textbf{0.722} \\
\textbf{Apache} &
  \textbf{1.000} &
  0.727 &
  \textbf{1.000} &
  \multicolumn{1}{l|}{0.517} &
  \textbf{1.000} &
  \textit{0.371} &
  \textbf{1.000} &
  \multicolumn{1}{l|}{\textit{0.448}} &
  \textit{0.976} &
  0.844 &
  \textit{0.949} &
  \multicolumn{1}{l|}{0.712} &
  \textbf{1.000} &
  \textbf{0.994} &
  \textbf{1.000} &
  \textbf{0.828} \\
\textbf{Zookeeper} &
  \textbf{0.994} &
  \textbf{0.844} &
  \textbf{0.904} &
  \multicolumn{1}{l|}{0.627} &
  0.993 &
  \textit{0.365} &
  0.885 &
  \multicolumn{1}{l|}{\textit{0.315}} &
  \textbf{0.994} &
  0.824 &
  \textbf{0.904} &
  \multicolumn{1}{l|}{0.723} &
  \textit{0.970} &
  0.825 &
  \textit{0.873} &
  \textbf{0.727} \\
\textbf{Mac} &
  \textit{0.761} &
  0.381 &
  \textit{0.229} &
  \multicolumn{1}{l|}{\textit{0.070}} &
  0.829 &
  \textit{0.206} &
  0.782 &
  \multicolumn{1}{l|}{0.137} &
  0.848 &
  0.539 &
  0.818 &
  \multicolumn{1}{l|}{0.370} &
  \textbf{0.871} &
  \textbf{0.553} &
  \textbf{0.883} &
  \textbf{0.551} \\
\textbf{Hadoop} &
  0.921 &
  0.541 &
  0.785 &
  \multicolumn{1}{l|}{0.394} &
  \textit{0.871} &
  \textit{0.385} &
  \textit{0.432} &
  \multicolumn{1}{l|}{\textit{0.151}} &
  \textbf{0.940} &
  0.812 &
  \textbf{0.947} &
  \multicolumn{1}{l|}{\textbf{0.598}} &
  0.908 &
  \textbf{0.693} &
  0.880 &
  0.574 \\
\textbf{OpenStack} &
  0.752 &
  \textit{0.020} &
  \textit{0.007} &
  \multicolumn{1}{l|}{\textit{0.002}} &
  \textbf{1.000} &
  0.189 &
  \textbf{1.000} &
  \multicolumn{1}{l|}{0.375} &
  \textit{0.316} &
  0.133 &
  0.279 &
  \multicolumn{1}{l|}{0.151} &
  0.950 &
  \textbf{0.417} &
  0.946 &
  \textbf{0.710} \\
\textbf{HealthApp} &
  0.862 &
  0.312 &
  \textit{0.010} &
  \multicolumn{1}{l|}{\textit{0.004}} &
  \textit{0.690} &
  \textit{0.011} &
  0.474 &
  \multicolumn{1}{l|}{0.034} &
  0.993 &
  0.656 &
  \textbf{0.946} &
  \multicolumn{1}{l|}{0.597} &
  \textbf{0.996} &
  \textbf{0.971} &
  0.914 &
  \textbf{0.763} \\
\textbf{HPC} &
  \textit{0.793} &
  0.721 &
  \textit{0.309} &
  \multicolumn{1}{l|}{0.147} &
  0.863 &
  \textit{0.098} &
  0.558 &
  \multicolumn{1}{l|}{\textit{0.084}} &
  \textit{0.793} &
  0.852 &
  0.584 &
  \multicolumn{1}{l|}{0.565} &
  \textbf{0.866} &
  \textbf{0.990} &
  \textbf{0.896} &
  \textbf{0.883} \\
\textbf{OpenSSH} &
  0.707 &
  0.586 &
  0.872 &
  \multicolumn{1}{l|}{0.487} &
  \textbf{0.780} &
  \textit{0.556} &
  \textbf{0.911} &
  \multicolumn{1}{l|}{\textit{0.456}} &
  0.707 &
  \textbf{0.649} &
  0.883 &
  \multicolumn{1}{l|}{0.857} &
  \textit{0.691} &
  0.633 &
  \textbf{0.921} &
  \textbf{0.868} \\
\textbf{BGL} &
  0.919 &
  \textit{0.456} &
  \textit{0.624} &
  \multicolumn{1}{l|}{\textit{0.204}} &
  0.925 &
  \textit{0.455} &
  0.641 &
  \multicolumn{1}{l|}{0.206} &
  \textbf{0.927} &
  0.822 &
  0.759 &
  \multicolumn{1}{l|}{0.588} &
  \textit{0.903} &
  \textbf{0.938} &
  \textbf{0.824} &
  \textbf{0.703} \\
\textbf{HDFS} &
  0.999 &
  \textit{0.569} &
  \textbf{0.935} &
  \multicolumn{1}{l|}{0.478} &
  0.999 &
  \textbf{0.948} &
  \textit{0.831} &
  \multicolumn{1}{l|}{\textbf{0.674}} &
  \textit{0.949} &
  0.720 &
  0.901 &
  \multicolumn{1}{l|}{\textit{0.418}} &
  \textbf{1.000} &
  0.721 &
  0.854 &
  0.449 \\
\textbf{Spark} &
  0.888 &
  0.559 &
  0.861 &
  \multicolumn{1}{l|}{0.448} &
  \textit{0.839} &
  \textit{0.280} &
  \textit{0.743} &
  \multicolumn{1}{l|}{\textit{0.253}} &
  0.889 &
  0.631 &
  0.866 &
  \multicolumn{1}{l|}{0.540} &
  \textbf{0.975} &
  \textbf{0.908} &
  \textbf{0.920} &
  \textbf{0.544} \\
\textbf{Thunderbird} &
  0.831 &
  0.219 &
  \textit{0.237} &
  \multicolumn{1}{l|}{\textit{0.072}} &
  \textit{0.766} &
  \textit{0.206} &
  0.241 &
  \multicolumn{1}{l|}{0.074} &
  0.829 &
  0.456 &
  \textbf{0.844} &
  \multicolumn{1}{l|}{0.472} &
  \textbf{0.846} &
  \textbf{0.556} &
  0.830 &
  \textbf{0.531} \\ \hline
\textbf{Average} &
  0.843 &
  0.481 &
  \textit{0.554} &
  \multicolumn{1}{l|}{\textit{0.278}} &
  0.858 &
  \textit{0.348} &
  0.706 &
  \multicolumn{1}{l|}{0.284} &
  0.847 &
  0.646 &
  0.815 &
  \multicolumn{1}{l|}{0.568} &
  \textbf{0.917} &
  \textbf{0.780} &
  \textbf{0.904} &
  \textbf{0.697} \\
\textbf{Gap w/ PIPLUP} &
  -8.1\% &
  -38.3\% &
  -38.7\% &
  \multicolumn{1}{l|}{-60.2\%} &
  -6.5\% &
  -55.4\% &
  -22.0\% &
  \multicolumn{1}{l|}{-59.3\%} &
  -7.7\% &
  -17.2\% &
  -9.9\% &
  \multicolumn{1}{l|}{-18.5\%} &
  - &
  - &
  - &
  - \\
\textbf{S-K ESD} &
  4 &
  6 &
  6 &
  \multicolumn{1}{l|}{6} &
  4 &
  7 &
  5 &
  \multicolumn{1}{l|}{6} &
  4 &
  4 &
  3 &
  \multicolumn{1}{l|}{4} &
  2 &
  3 &
  1 &
  2 \\ \hline
\multirow{3}{*}{\textbf{Dataset}} &
  \multicolumn{16}{c}{\textbf{Semantic-based Parsers}} \\ \cline{2-17} 
 &
  \multicolumn{4}{c|}{\textbf{LILAC (GPT-3.5)}} &
  \multicolumn{4}{c|}{\textbf{LibreLog (Llama3-8B)}} &
  \multicolumn{4}{c|}{\textbf{LogBatcher (GPT-3.5)}} &
  \multicolumn{4}{c}{\textbf{LUNAR (GPT-3.5)}} \\ \cline{2-17} 
 &
  \multicolumn{1}{c}{\textbf{GA}} &
  \multicolumn{1}{c}{\textbf{PA}} &
  \multicolumn{1}{c}{\textbf{FGA}} &
  \multicolumn{1}{c|}{\textbf{FTA}} &
  \multicolumn{1}{c}{\textbf{GA}} &
  \multicolumn{1}{c}{\textbf{PA}} &
  \multicolumn{1}{c}{\textbf{FGA}} &
  \multicolumn{1}{c|}{\textbf{FTA}} &
  \multicolumn{1}{c}{\textbf{GA}} &
  \multicolumn{1}{c}{\textbf{PA}} &
  \multicolumn{1}{c}{\textbf{FGA}} &
  \multicolumn{1}{c|}{\textbf{FTA}} &
  \multicolumn{1}{c}{\textbf{GA}} &
  \multicolumn{1}{c}{\textbf{PA}} &
  \multicolumn{1}{c}{\textbf{FGA}} &
  \multicolumn{1}{c}{\textbf{FTA}} \\ \hline
\textbf{Proxifier} &
  \textit{0.000} &
  \textit{0.099} &
  \textit{0.000} &
  \multicolumn{1}{l|}{\textit{0.065}} &
  0.510 &
  0.897 &
  0.500 &
  \multicolumn{1}{l|}{0.583} &
  0.676 &
  0.676 &
  0.727 &
  \multicolumn{1}{l|}{0.727} &
  0.989 &
  \textbf{1.000} &
  0.870 &
  0.957 \\
\textbf{Linux} &
  0.942 &
  0.729 &
  0.819 &
  \multicolumn{1}{l|}{\textit{0.447}} &
  0.967 &
  0.846 &
  \textit{0.809} &
  \multicolumn{1}{l|}{0.613} &
  \textbf{0.946} &
  \textbf{0.906} &
  \textbf{0.905} &
  \multicolumn{1}{l|}{0.679} &
  \textit{0.831} &
  \textit{0.741} &
  0.862 &
  \textbf{0.720} \\
\textbf{Apache} &
  \textbf{1.000} &
  0.970 &
  \textbf{1.000} &
  \multicolumn{1}{l|}{\textit{0.621}} &
  \textbf{1.000} &
  0.993 &
  \textbf{1.000} &
  \multicolumn{1}{l|}{0.759} &
  \textit{0.997} &
  0.990 &
  \textit{0.921} &
  \multicolumn{1}{l|}{0.698} &
  \textbf{1.000} &
  \textbf{0.998} &
  \textbf{1.000} &
  \textbf{0.862} \\
\textbf{Zookeeper} &
  \textit{0.987} &
  \textit{0.375} &
  \textit{0.867} &
  \multicolumn{1}{l|}{\textit{0.578}} &
  \textbf{1.000} &
  0.850 &
  \textbf{0.980} &
  \multicolumn{1}{l|}{\textbf{0.863}} &
  0.990 &
  0.825 &
  0.914 &
  \multicolumn{1}{l|}{0.823} &
  \textbf{0.993} &
  \textbf{0.851} &
  0.885 &
  0.800 \\
\textbf{Mac} &
  \textit{0.707} &
  \textit{0.405} &
  \textit{0.770} &
  \multicolumn{1}{l|}{\textit{0.393}} &
  0.814 &
  \textbf{0.651} &
  0.824 &
  \multicolumn{1}{l|}{0.498} &
  0.856 &
  0.520 &
  0.863 &
  \multicolumn{1}{l|}{0.491} &
  \textbf{0.882} &
  0.605 &
  \textbf{0.869} &
  \textbf{0.551} \\
\textbf{Hadoop} &
  \textit{0.890} &
  0.711 &
  0.921 &
  \multicolumn{1}{l|}{\textit{0.640}} &
  \textbf{0.964} &
  \textbf{0.865} &
  \textbf{0.946} &
  \multicolumn{1}{l|}{\textbf{0.754}} &
  0.923 &
  \textit{0.634} &
  \textit{0.862} &
  \multicolumn{1}{l|}{0.651} &
  0.941 &
  0.840 &
  0.928 &
  0.687 \\
\textbf{OpenStack} &
  \textbf{1.000} &
  0.477 &
  \textbf{1.000} &
  \multicolumn{1}{l|}{0.792} &
  0.811 &
  0.831 &
  \textit{0.667} &
  \multicolumn{1}{l|}{\textit{0.624}} &
  \textit{0.524} &
  \textit{0.447} &
  0.957 &
  \multicolumn{1}{l|}{0.681} &
  \textbf{1.000} &
  \textbf{0.952} &
  \textbf{1.000} &
  \textbf{0.875} \\
\textbf{HealthApp} &
  0.993 &
  \textit{0.542} &
  0.971 &
  \multicolumn{1}{l|}{\textit{0.711}} &
  \textit{0.992} &
  0.838 &
  \textbf{0.975} &
  \multicolumn{1}{l|}{\textbf{0.862}} &
  0.998 &
  \textbf{0.976} &
  \textit{0.935} &
  \multicolumn{1}{l|}{0.806} &
  \textbf{1.000} &
  0.963 &
  0.971 &
  0.850 \\
\textbf{HPC} &
  0.870 &
  \textit{0.691} &
  0.857 &
  \multicolumn{1}{l|}{\textit{0.586}} &
  \textbf{0.967} &
  0.846 &
  \textit{0.838} &
  \multicolumn{1}{l|}{0.662} &
  \textit{0.861} &
  0.985 &
  \textbf{0.901} &
  \multicolumn{1}{l|}{\textbf{0.861}} &
  0.864 &
  \textbf{0.990} &
  0.828 &
  0.817 \\
\textbf{OpenSSH} &
  0.745 &
  \textit{0.347} &
  \textit{0.727} &
  \multicolumn{1}{l|}{\textit{0.364}} &
  \textbf{0.868} &
  0.496 &
  0.806 &
  \multicolumn{1}{l|}{0.472} &
  \textit{0.726} &
  \textbf{0.925} &
  0.840 &
  \multicolumn{1}{l|}{0.765} &
  0.780 &
  0.722 &
  \textbf{0.923} &
  \textbf{0.923} \\
\textbf{BGL} &
  \textit{0.833} &
  \textit{0.745} &
  0.820 &
  \multicolumn{1}{l|}{\textit{0.582}} &
  0.907 &
  0.929 &
  \textit{0.808} &
  \multicolumn{1}{l|}{0.722} &
  0.941 &
  0.881 &
  0.817 &
  \multicolumn{1}{l|}{0.689} &
  \textbf{0.957} &
  \textbf{0.968} &
  \textbf{0.885} &
  \textbf{0.807} \\
\textbf{HDFS} &
  \textbf{1.000} &
  \textit{0.909} &
  \textit{0.658} &
  \multicolumn{1}{l|}{\textit{0.493}} &
  \textbf{1.000} &
  \textbf{1.000} &
  0.886 &
  \multicolumn{1}{l|}{0.795} &
  \textbf{1.000} &
  0.999 &
  \textbf{1.000} &
  \multicolumn{1}{l|}{0.913} &
  \textbf{1.000} &
  \textbf{1.000} &
  0.968 &
  \textbf{0.968} \\
\textbf{Spark} &
  \textbf{0.999} &
  0.688 &
  0.885 &
  \multicolumn{1}{l|}{0.568} &
  \textit{0.859} &
  0.889 &
  0.865 &
  \multicolumn{1}{l|}{\textbf{0.712}} &
  0.943 &
  \textit{0.715} &
  \textit{0.788} &
  \multicolumn{1}{l|}{\textit{0.593}} &
  0.974 &
  \textbf{0.996} &
  \textbf{0.880} &
  \textit{0.656} \\
\textbf{Thunderbird} &
  \textit{0.757} &
  \textit{0.491} &
  \textit{0.322} &
  \multicolumn{1}{l|}{\textit{0.177}} &
  \textbf{0.874} &
  \textbf{0.689} &
  0.862 &
  \multicolumn{1}{l|}{0.574} &
  0.750 &
  0.556 &
  0.813 &
  \multicolumn{1}{l|}{0.525} &
  0.867 &
  0.639 &
  \textbf{0.878} &
  \textbf{0.593} \\ \hline
\textbf{Average} &
  \textit{0.837} &
  \textit{0.584} &
  \textit{0.758} &
  \multicolumn{1}{l|}{\textit{0.501}} &
  0.895 &
  0.830 &
  0.840 &
  \multicolumn{1}{l|}{0.678} &
  0.866 &
  0.788 &
  0.874 &
  \multicolumn{1}{l|}{0.707} &
  \textbf{0.934} &
  \textbf{0.876} &
  \textbf{0.911} &
  \textbf{0.790} \\
\textbf{Gap w/ PIPLUP} &
  -8.7\% &
  -25.1\% &
  -16.2\% &
  \multicolumn{1}{l|}{-28.1\%} &
  -2.4\% &
  +6.4\% &
  -7.1\% &
  \multicolumn{1}{l|}{-2.8\%} &
  -5.5\% &
  +1.0\% &
  -3.3\% &
  \multicolumn{1}{l|}{+1.4\%} &
  +1.8\% &
  +12.3\% &
  +0.7\% &
  +13.4\% \\
\textbf{S-K ESD} &
  4 &
  5 &
  4 &
  \multicolumn{1}{l|}{5} &
  3 &
  2 &
  3 &
  \multicolumn{1}{l|}{3} &
  4 &
  3 &
  2 &
  \multicolumn{1}{l|}{2} &
  1 &
  1 &
  1 &
  1 \\ \hline
\end{tabular}%
}
\end{table}
 The default parameters are used to parse all 14 datasets. Table~\ref{tab:effectiveness} shows the parsing effectiveness of PIPLUP for all 14 studied datasets, along with the eight baselines. Overall, PIPLUP statistically outperforms the state-of-the-art statistic-based benchmarks. Even when considering semantic-based parsers in the comparison, PIPLUP's performance is consistently optimal or near-optimal: it achieved the highest GA among all tools, and has the same level of performance as LogBatcher on PA and FTA.

\textbf{PIPLUP obtains significantly better average performance than state-of-the-art statistic-based parsers on all four metrics. }
As a result of several novel technical considerations (e.g., dynamically searching for constant tokens and configuring length-based similarity thresholds for log grouping, using a frequency-based branching method to isolate highly similar but semantically different templates), PIPLUP achieves high and robust parsing performance. 
According to the Scott-Knott ESD ranking, PIPLUP consistently achieves a higher ranking than all the statistic-based parsers. In comparison, existing statistic-based approaches heavily depend on some ungeneralizable heuristics. For example, Drain assumes that constant tokens exist in the prefix. The statement turns out to be not highly effective according to Table~\ref{tab:effectiveness}: Drain achieved one of the lowest average values on GA and FGA among the tools, suggesting unstable performance; even with the shuffling strategy, XDrain failed to overcome the challenge. Conversely, our dynamic constant token searching and template merging strategy can effectively cluster log messages that indicate the same events. 

\textbf{The data-insensitive parameter setting of PIPLUP is robust on new datasets.} As shown in Table~\ref{tab:effectiveness}, PIPLUP achieves robust performance across all 14 datasets, including the four largest datasets (Spark, Thunderbird, BGL, and HDFS) not used in determining the default parameters in RQ1. For the four largest log datasets, PIPLUP achieves an average GA, PA, FGA, and FTA of 0.931, 0.781, 0.857, and 0.557, respectively, outperforming all the statistic-based parsers. 
Apart from better explainability and fewer configuration efforts, the length-dependent $\theta_{sim}$ tends to be more generalizable than the data-sensitive thresholds used in Drain and its variants, as they can be directly deployed on other datasets without the need for adequate prior knowledge. The branching allowance (i.e., $\theta_{br}$) also largely enhanced the parsing performance on Apache and Spark: shown as the successful case in Table~\ref{tab:branch_cases}, the two different events are correctly separated by PIPLUP's branching stage, although they are highly similar. On the other hand, branching may also slightly decrease the accuracy due to the small number of choices in variables, illustrated as the failed case in Table~\ref{tab:branch_cases}. The two events should belong to the same cluster, as their last token is a variable. However, since the log file only contains these two choices for this variable, PIPLUP will identify them as constants under the default settings of $\theta_{br}=2$. We argue that the issue would be suppressed when adequate message samples are presented to PIPLUP. 

We noticed that PIPLUP suffered from performance degradation on HDFS in terms of PA and FTA, failing to exceed XDrain. The reason roots in the preprocessing processes. During the preprocessing stage, the class names starting with ``java.io'' are automatically identified as variables. Practitioners can suppress this problem by refining the preprocessing regexes with their domain knowledge. 

\textbf{PIPLUP achieves better results than most LLM-based log parsers, except LUNAR.}
As shown in Table~\ref{tab:effectiveness}, PIPLUP achieves the 1st ranking in FGA and 2nd ranking in GA, outperforming three out of four semantic-based parsers, suggesting its robust performance in log grouping. It is statistically on the same performance level as LogBatcher in terms of PA and FTA, suggesting its competitive performance in parsing accuracy. 

It can be observed that while the performance gaps on grouping metrics are small (1.8\% for GA and 0.7\% for FGA), LUNAR obtained more than 13\% improvements on FTA in comparison to PIPLUP, and is statistically the best unsupervised log parser. The results are expected, as GPT-3.5 is trained on a large corpus; the external knowledge from both its training data and the prompt allows LUNAR to precisely identify the variables and infer templates. However, using commercial LLMs such as GPT-3.5 can impose significant costs and privacy concerns~\citep{ekhlasi2025insightai,sallou2024breaking}. According to a previous study by \citet{aghili2025protecting}, industrial practitioners regard IP addresses, MAC addresses, host names, and file paths as the most sensitive information in logs. Although IP addresses and MAC addresses can be easily identified using regexes~\citep{qin2025preprocessing}, file paths and host names are sometimes difficult to mask due to their heterogeneity in patterns. Considering the risk of information leaks, commercial LLMs for log analysis are less favorable for organizations.

\begin{boxD}
PIPLUP is effective and robust for log parsing, outperforming state-of-the-art statistic-based log parsers on all accuracy metrics. Even with LLM-powered semantic-based parsers included, the simple PIPLUP approach is still competitive in terms of all four metrics.
\end{boxD}

\subsection*{\textbf{RQ3: How does PIPLUP compare to state-of-the-art parsers in terms of parsing efficiency?}}
\label{sec:rq3}

We use the default parameter to parse all datasets. Due to resource limitations, we did not replicate 
LibreLog~\citep{ma2024librelog}, as we use the same datasets as used in their original studies and their deployment environment is known. Since LILAC, LogBatcher, and LUNAR call the OpenAI API~\citep{ChatGPT}, the time consumption reflects their sampling, caching, and server response time. The datasets are sorted from the smallest (i.e., with the least number of lines) to the largest (i.e., with the most number of lines), and their time consumptions are documented in Table~\ref{tab:efficiency}. 

\begin{table}[]
\centering
\caption{The time consumption of log parsers on different datasets. }
\label{tab:efficiency}
\resizebox{\columnwidth}{!}{%
\begin{tabular}{l|l|l|llllllll}
\hline
\multirow{3}{*}{\textbf{Dataset}} &
  \multicolumn{1}{c|}{\multirow{3}{*}{\textbf{\# Lines}}} &
  \multirow{3}{*}{\textbf{\# Temp}} &
  \multicolumn{8}{c}{\textbf{Time Consumption (s)}} \\ \cline{4-11} 
 &
  \multicolumn{1}{c|}{} &
   &
  \multicolumn{4}{c|}{\textbf{Statistic-based Parsers}} &
  \multicolumn{4}{c}{\textbf{Semantic-based Parsers}} \\ \cline{4-11} 
 &
  \multicolumn{1}{c|}{} &
   &
  \multicolumn{1}{c|}{\textbf{Drain}} &
  \multicolumn{1}{l|}{\textbf{XDrain}} &
  \multicolumn{1}{c|}{\textbf{Pre-Drain}} &
  \multicolumn{1}{c|}{\textbf{PIPLUP}} &
  \multicolumn{1}{c|}{\textbf{LILAC}} &
  \multicolumn{1}{c|}{\textbf{LibreLog}} &
  \multicolumn{1}{l|}{\textbf{LogBatcher}} &
  \textbf{LUNAR} \\ \hline
\textbf{Proxifier} &
  21,320 &
  11 &
  \multicolumn{1}{l|}{1.3} &
  \multicolumn{1}{l|}{3.1} &
  \multicolumn{1}{l|}{1.1} &
  \multicolumn{1}{l|}{1.2} &
  \multicolumn{1}{l|}{20.2} &
  \multicolumn{1}{l|}{871.5} &
  \multicolumn{1}{l|}{15.9} &
  26.8 \\
\textbf{Linux} &
  23,921 &
  338 &
  \multicolumn{1}{l|}{1.1} &
  \multicolumn{1}{l|}{3.1} &
  \multicolumn{1}{l|}{1.4} &
  \multicolumn{1}{l|}{2.3} &
  \multicolumn{1}{l|}{930.3} &
  \multicolumn{1}{l|}{216.0} &
  \multicolumn{1}{l|}{265.4} &
  301.7 \\
\textbf{Apache} &
  51,977 &
  29 &
  \multicolumn{1}{l|}{1.7} &
  \multicolumn{1}{l|}{3.7} &
  \multicolumn{1}{l|}{2.1} &
  \multicolumn{1}{l|}{2.2} &
  \multicolumn{1}{l|}{43.2} &
  \multicolumn{1}{l|}{18.9} &
  \multicolumn{1}{l|}{22.0} &
  34.5 \\
\textbf{Zookeeper} &
  74,273 &
  89 &
  \multicolumn{1}{l|}{2.4} &
  \multicolumn{1}{l|}{9.6} &
  \multicolumn{1}{l|}{3.2} &
  \multicolumn{1}{l|}{3.5} &
  \multicolumn{1}{l|}{74.1} &
  \multicolumn{1}{l|}{52.2} &
  \multicolumn{1}{l|}{58.7} &
  81.0 \\
\textbf{Mac} &
  100,314 &
  626 &
  \multicolumn{1}{l|}{4.5} &
  \multicolumn{1}{l|}{18.8} &
  \multicolumn{1}{l|}{6.8} &
  \multicolumn{1}{l|}{9.7} &
  \multicolumn{1}{l|}{747.5} &
  \multicolumn{1}{l|}{5,935.8} &
  \multicolumn{1}{l|}{642.4} &
  17311.0 \\
\textbf{Hadoop} &
  179,993 &
  236 &
  \multicolumn{1}{l|}{5.5} &
  \multicolumn{1}{l|}{30.7} &
  \multicolumn{1}{l|}{10.2} &
  \multicolumn{1}{l|}{11.5} &
  \multicolumn{1}{l|}{175.7} &
  \multicolumn{1}{l|}{285.7} &
  \multicolumn{1}{l|}{914.3} &
  334.9 \\
\textbf{OpenStack} &
  207,632 &
  48 &
  \multicolumn{1}{l|}{23.8} &
  \multicolumn{1}{l|}{22.0} &
  \multicolumn{1}{l|}{14.1} &
  \multicolumn{1}{l|}{13.8} &
  \multicolumn{1}{l|}{54.5} &
  \multicolumn{1}{l|}{377.6} &
  \multicolumn{1}{l|}{40.0} &
  698.6 \\
\textbf{HealthApp} &
  212,394 &
  156 &
  \multicolumn{1}{l|}{5.1} &
  \multicolumn{1}{l|}{23.5} &
  \multicolumn{1}{l|}{6.3} &
  \multicolumn{1}{l|}{6.7} &
  \multicolumn{1}{l|}{107.6} &
  \multicolumn{1}{l|}{103.3} &
  \multicolumn{1}{l|}{100.3} &
  153.0 \\
\textbf{HPC} &
  429,987 &
  74 &
  \multicolumn{1}{l|}{11.9} &
  \multicolumn{1}{l|}{47.4} &
  \multicolumn{1}{l|}{11.4} &
  \multicolumn{1}{l|}{11.8} &
  \multicolumn{1}{l|}{47.5} &
  \multicolumn{1}{l|}{539.8} &
  \multicolumn{1}{l|}{79.1} &
  177.0 \\
\textbf{OpenSSH} &
  638,946 &
  38 &
  \multicolumn{1}{l|}{20.6} &
  \multicolumn{1}{l|}{19.8} &
  \multicolumn{1}{l|}{21.0} &
  \multicolumn{1}{l|}{24.0} &
  \multicolumn{1}{l|}{43.1} &
  \multicolumn{1}{l|}{89.4} &
  \multicolumn{1}{l|}{76.6} &
  129.8 \\
\textbf{BGL} &
  4,631,261 &
  320 &
  \multicolumn{1}{l|}{147.0} &
  \multicolumn{1}{l|}{148.8} &
  \multicolumn{1}{l|}{151.8} &
  \multicolumn{1}{l|}{149.2} &
  \multicolumn{1}{l|}{422.4} &
  \multicolumn{1}{l|}{1,244.6} &
  \multicolumn{1}{l|}{429.3} &
  872.3 \\
\textbf{HDFS} &
  11,167,740 &
  46 &
  \multicolumn{1}{l|}{361.9} &
  \multicolumn{1}{l|}{426.6} &
  \multicolumn{1}{l|}{533.0} &
  \multicolumn{1}{l|}{576.5} &
  \multicolumn{1}{l|}{418.7} &
  \multicolumn{1}{l|}{1,252.6} &
  \multicolumn{1}{l|}{96.8} &
  1278.1 \\
\textbf{Spark} &
  16,075,117 &
  236 &
  \multicolumn{1}{l|}{554.4} &
  \multicolumn{1}{l|}{2,497.5} &
  \multicolumn{1}{l|}{747.5} &
  \multicolumn{1}{l|}{902.3} &
  \multicolumn{1}{l|}{625.1} &
  \multicolumn{1}{l|}{1,752.4} &
  \multicolumn{1}{l|}{286.5} &
  2148.9 \\
\textbf{Thunderbird} &
  16,601,745 &
  1,241 &
  \multicolumn{1}{l|}{1,110.0} &
  \multicolumn{1}{l|}{1,207.4} &
  \multicolumn{1}{l|}{6,846.3} &
  \multicolumn{1}{l|}{1445.3} &
  \multicolumn{1}{l|}{8000.4} &
  \multicolumn{1}{l|}{8,659.3} &
  \multicolumn{1}{l|}{10375.6} &
  25917.5 \\ \hline
\textbf{Average} &
  3,601,187 &
  249.1 &
  \multicolumn{1}{l|}{160.8} &
  \multicolumn{1}{l|}{318.7} &
  \multicolumn{1}{l|}{596.9} &
  \multicolumn{1}{l|}{225.7} &
  \multicolumn{1}{l|}{836.4} &
  \multicolumn{1}{l|}{1,528.5} &
  \multicolumn{1}{l|}{916.3} &
  3533.2 \\
\textbf{S-K ESD} &
  - &
  - &
  \multicolumn{1}{l|}{1} &
  \multicolumn{1}{l|}{2} &
  \multicolumn{1}{l|}{3} &
  \multicolumn{1}{l|}{1} &
  \multicolumn{1}{l|}{3} &
  \multicolumn{1}{l|}{4} &
  \multicolumn{1}{l|}{2} &
  5 \\
\textbf{S-K ESD w/o Thunderbird} &
  - &
  - &
  \multicolumn{1}{l|}{1} &
  \multicolumn{1}{l|}{2} &
  \multicolumn{1}{l|}{1} &
  \multicolumn{1}{l|}{1} &
  \multicolumn{1}{l|}{3} &
  \multicolumn{1}{l|}{4} &
  \multicolumn{1}{l|}{2} &
  5 \\ \hline
\end{tabular}%
}
\end{table}

\textbf{The parsing time not only depends on the size of logs, but also on the complexity of logs (e.g., diversity of templates).} 
As shown in Table~\ref{tab:efficiency}, in general, for all log parsers, including PIPLUP, a larger log file takes a longer time to parse, with some exceptions.
In particular, it can be observed that all tools exhibited abnormal time consumption patterns when profiling the largest log file, Thunderbird, which also has the highest template variety. Log parsers often rely on the pattern-matching stage to find the most suitable event cluster for an incoming log message, and this stage may lead to high time consumption when the number of pattern candidates (i.e., the number of different templates) is large. 

\textbf{PIPLUP has a comparable time consumption to Drain.} As shown in Table~\ref{tab:efficiency}, in general, all four statistic-based parsers exhibited relatively low time consumption in parsing most of the log files. 
According to the Scott-Knott ESD ranking, PIPLUP is consistently ranked at 1st place as Drain, showing that it is statistically faster than other tools. 
Both Drain and Preprocessed-Drain infer only one template for each cluster and do not require the template matching process described in Sec.~\ref{sec:template_matching}, thus their time consumption is lower than PIPLUP on most datasets. However, we noticed that Preprocessed-Drain suffered from high time consumption on Thunderbird from Table~\ref{tab:efficiency}, since the change in variable masking during preprocessing would change the behavior of Drain's template extraction components~\citep{qin2025preprocessing}. 
XDrain's time consumption is correlated to the number of times logs are shuffled, since each shuffle indicates a re-run of Drain. For example, the default number of shuffles for Spark is set to 5, and thus, XDrain's parsing time on this dataset is around five times that of Drain's. 


\textbf{PIPLUP has a much lower and predictable execution time than semantic-based parsers.} Although effective caching and parsing approaches are applied, all four semantic-based parsers demand more time to parse the 14 datasets on average, in comparison to PIPLUP. Further, we noticed that all semantic-based parsers demand more than 500 seconds when parsing the Mac dataset: a small dataset with highly diverse event templates. 
The abnormal parsing time on small datasets may hinder the semantic-based parsers' usage in real-world scenarios, given that their time consumption is highly non-deterministic. In comparison with our approach, the average time for PIPLUP to parse all datasets is 26.985\% of LILAC's time consumption, 14.766\% of LibreLog's, 24.632\% of LogBatcher's, and 6.388\% of LUNAR's. The result suggests that PIPLUP can more efficiently parse logs than semantic-based parsers, even if powerful machines with 
GPU acceleration is involved. 
Besides, according to Table~\ref{tab:efficiency}, PIPLUP shows a much more stable and lower time consumption increment trend. 

\begin{boxD}
PIPLUP is efficient at log parsing, requiring only $\sim$1.5 seconds to $\sim$25 minutes to parse each of the studied datasets. Its time efficiency is statistically comparable to state-of-the-art statistic-based parsers and much better than semantic-based ones that rely on expensive computing resources (e.g., only using ~6\% of LUNAR's parsing time). 
\end{boxD}

\section{Discussions}
\label{sec:discussions}
In the discussion section, we systematically compare the different parsing approaches in terms of parsing effectiveness, efficiency, required computational resources, and whether they raise privacy issues. The comparison is based on the Scott-Knott effect size difference rankings in Table~\ref{tab:effectiveness} and \ref{tab:efficiency} and their implementation details. Based on the comparison, we make recommendations to researchers and practitioners on which parser to select according to the application context. 

\begin{table}[]
\centering
\caption{A qualitative comparison of parser effectiveness, efficiency, required computational resources, and privacy concerns. The high/low are classified with the Scott-Knott effect size difference rankings. }
\label{tab:comparisons}
\resizebox{\columnwidth}{!}{%
\begin{tabular}{l|llll|llll}
\hline
\multicolumn{1}{c|}{\textbf{}} & \multicolumn{4}{c|}{\textbf{Statistic-based Parsers}} & \multicolumn{4}{c}{\textbf{Semantic-based Parsers}} \\ \hline
\multicolumn{1}{c|}{\textbf{}} &
  \multicolumn{1}{c}{\textbf{Drain}} &
  \multicolumn{1}{c}{\textbf{XDrain}} &
  \multicolumn{1}{c}{\textbf{Pre-Drain}} &
  \multicolumn{1}{c|}{\textbf{PIPLUP}} &
  \multicolumn{1}{c}{\textbf{LILAC}} &
  \multicolumn{1}{c}{\textbf{LibreLog}} &
  \multicolumn{1}{c}{\textbf{LogBatcher}} &
  \multicolumn{1}{c}{\textbf{LUNAR}} \\ \hline
\textbf{GA}                    & Low        & Low        & Low           & High        & Low         & Low         & Low        & High       \\ \hline
\textbf{PA}                    & Low        & Low        & Medium        & High        & Low         & High        & High       & High       \\ \hline
\textbf{FGA}                   & Low        & Low        & Low           & High        & Low         & High        & High       & High       \\ \hline
\textbf{FTA}                   & Low        & Low        & Low           & High        & Low         & High        & High       & High       \\ \hline
\textbf{Time}                  & Low        & Low        & Medium        & Low         & High        & High        & Low        & High       \\ \hline
\textbf{Comp. Resource}        & CPU         & CPU         & CPU            & CPU          & API Credits         & GPU          & API Credits        & API Credits        \\ \hline
\textbf{Privacy Issues}        & No         & No         & No            & No          & Yes         & No          & Yes        & Yes        \\ \hline
\end{tabular}%
}
\end{table}

As shown in Table~\ref{tab:comparisons}, in comparison to the remaining four statistic-based parsers, PIPLUP reached high performance in terms of all four performance metrics and maintained a low time requirement. It also does not raise privacy issues, since its execution does not require any closed-source models. On the other hand, among the four semantic-based parsers, only LUNAR reached a high overall parsing performance. However, it requires high time consumption and may raise privacy concerns due to the usage of GPT-3.5. In exchange for a lower grouping accuracy, LibreLog prevents potential privacy issues by using Llama3 locally, and LogBatcher can significantly reduce the parsing time. 

The major applications of log parsers are industrial use and log-related academic use. Given that privacy and cost are of major concerns in the industrial scenario, we argue that PIPLUP is more preferable than the semantic-based parsers. Although LibreLog also ensures information privacy, its high time consumption, even with GPU acceleration, hinders its application, since industrial systems always produce large volumes of log lines. 

In academia, researchers often use log parsers to support their downstream research, such as anomaly detection and root cause analysis. Therefore, the quality of the parsing results becomes the major concern, because inaccurate results may lead to issues in downstream analyses. Given that typically a small number of benchmark datasets are analyzed, we think that LUNAR might be the best choice to summarize high-quality log templates with the aid of GPT-3.5 or other LLMs such as GPT-4o. In comparison, PIPLUP could be a suboptimal choice when computational resources (i.e., API credit) or time is limited for the research.

\section{Related Works}
\label{sec:related_works}

We discuss log parsing approaches along two categories: statistic-based and semantic-based, similar to prior work~\citep{jiang2024large}.

\subsection{Statistic-based Log Parsing}
\label{sec:background_statistic_based}
Statistic-based log parsers leverage statistical facts, such as frequency~\citep{dai2020logram, vaarandi2003data, nagappan2010abstracting, vaarandi2015logcluster}, similarity~\citep{fu2009execution, hamooni2016logmine, mizutani2013incremental, shima2016length, tang2011logsig}, and heuristic rules to group logs and summarize templates~\citep{jiang2008abstracting, du2016spell, he2017drain,makanju2009clustering, messaoudi2018search, yu2023brain,dai2020logram,dai2023pilar,liu2024xdrain}. These tools require no labels and fewer computational resources during log parsing, and are generally more time-effective than semantic-based parsers. According to a large-scale study by \citet{jiang2024large}, these log parsers are able to achieve high grouping accuracies. For example, the state-of-the-art statistic-based parser, Drain~\citep{he2017drain}, reached a high average grouping accuracy of 0.840 and only requires around 1.6 hours to parse all 14 datasets in the Loghub 2.0 dataset. However, due to the lack of external knowledge of variable semantics, Drain's parsing accuracies are limited in comparison to semantic-based approaches~\citep{jiang2024large}. To tackle this problem, recent research enhanced the parsers' preprocessing~\citep{qin2025preprocessing} and postprocessing~\citep{fu2022investigating} stages by inputting generalizable knowledge, and also improved their heuristics~\citep{liu2024xdrain}. Another point worth noticing is that most current approaches, such as Drain, requires data-sensitive parameters for log parsing, while it is often difficult to set optimal parameter values without adequate prior knowledge and experiments~\citep{dai2023pilar,chu2021prefix}. For better usability, we propose a data-insensitive log parser with a novel tree structure that dynamically identifies constant tokens, leverage dynamically-configured message-specific parameters and globally-insensitive parameters, and automatically merges similar event templates to prevent redundancy. Our design achieves effective, robust, and efficient log parsing. 

\subsection{Semantic-based Log Parsing}
\label{sec:background_semantic_based}
Semantic-based log parsers leverage the power of language models to learn the log patterns, identify variables, and generate log templates. 
This procedure can be done in either an unsupervised or a supervised fashion: supervised parsers such as UniParser~\citep{liu2022uniparser}, LogPPT~\citep{le2023log}, few-shot LILAC~\citep{jiang2024lilac}, and UNLEASH~\citep{le2025unleashing} are highly effective but require sample labels (i.e., log templates) provided by users to infer the templates for the remaining log messages; recent unsupervised semantic-based log parsers, LibreLog~\citep{ma2024librelog}, LogBatcher~\citep{xiao2024free}, and LUNAR~\citep{huang2025no} exempted the requirement of labels in parsing, and LILAC~\citep{jiang2024lilac} is also able to parse logs in an unsupervised method (i.e., through zero-shot learning). Semantic-based log parsers can obtain both high grouping accuracies and parsing accuracies, but often require expensive computational resources. Moreover, even advanced platforms will not ensure moderate time consumption of these parsers. For instance, even though LibreLog~\citep{ma2024librelog} considers parsing acceleration in its design (e.g., using log template memory search), requiring around $\frac{1}{3}$ time of LILAC, 
it still requires more than 3$\times$ the time consumption of the original Drain on the same amount of logs. Further, the usage of black-box models such as ChatGPT could raise privacy issues when the tool is applied in the industry~\citep{sallou2024breaking}. The high time complexity and deployment cost, and concerns toward privacy leak~\citep{aghili2025protecting} made semantic-based log parsers difficult to use for parsing daily logs produced in real-world scenarios. 
\section{Threats to Validity}
\label{sec:validity_threat}

\noindent\underline{\textbf{Internal Validity. }}
We aim to propose a log parser without configuration requirements, leveraging message-specific parameters and globally-insensitive parameters, for better usability. Nevertheless, PIPLUP has two inherent parameters whose default configurations may not achieve optimal performance for all datasets. To prove the generalizability of our settings, we left out the four largest datasets when determining the default settings. Our evaluation on the left-out datasets proves the generalizability of the default settings. 

\noindent\underline{\textbf{External Validity. }}
In this study, we used the 14 public log datasets provided in Loghub 2.0. The log files vary in size and come from different systems. 
However, PIPLUP's usability in other datasets, in particular, industrial scenarios is to be validated in future works.

\noindent\underline{\textbf{Construct Validity. }} 
Due to resource limitations, we did not replicate the results of LibreLog, but obtained its results from the original work. 
As LLM models are rapidly updated with online data, we acknowledge that ChatGPT and Llama3-8B may achieve higher effectiveness when trained with more knowledge related to log parsing. 

\section{Conclusions}
\label{sec:conclusion_future_work}
Our work proposes PIPLUP, a novel configuration-free statistic-based log parser, which is highly accurate, robust, and efficient. The technical contributions of PIPLUP include a novel tree structure without the fixed constant token assumptions, an enhanced template extraction process, and carefully chosen data-insensitive parameters.
The evaluation with 14 widely used log datasets shows that PIPLUP outperforms state-of-the-art statistic-based log parsers in terms of both accuracy and efficiency. Furthermore, the parsing effectiveness of our simple classic approach is close to the best LLM-powered unsupervised semantic-based parser (i.e., LUNAR), while being much faster (e.g., using less than one-tenth of LUNAR's time on average), resource efficient, and privacy-preserving.  
Our experiment result challenges the previous belief that semantic-based parsers are more effective than statistic-based parsers, encouraging future log analysis work to pay attention to classic approaches in this overwhelming LLM era, as simplicity, efficiency, and privacy may promote better adoption with enhanced usability. 

\balance
\bibliographystyle{plainnat}
\bibliography{references}

\end{document}